%% file: mainArXiv.tex
\documentclass[acmlarge, screen=true, nonacm]{acmart}

\input{00packages.tex}

\AtBeginDocument{%
  \providecommand\BibTeX{{%
    \normalfont B\kern-0.5em{\scshape i\kern-0.25em b}\kern-0.8em\TeX}}}

\begin{document}

\title[Mapping Caregiver Needs to AI Chatbot Design: Strengths and Gaps in Mental Health Support for AD/ADRD Caregivers]{Mapping Caregiver Needs to AI Chatbot Design: Strengths and Gaps in Mental Health Support for Alzheimer's and Dementia Caregivers}
\author{Jiayue Melissa Shi}
\orcid{0009-0007-0624-2421}
\affiliation{%
  \institution{University of Illinois Urbana-Champaign}
 \city{Urbana}
 \state{IL}
 \country{USA}}
 \email{mshi24@illinois.edu}

\author{Dong Whi Yoo}
\orcid{0000-0003-2738-1096}
\affiliation{%
 \institution{Indiana University Indianapolis}
 \city{Indianapolis}
 \state{IN}
 \country{USA}}
 \email{dy22@iu.edu}

 \author{Keran Wang}
\orcid{0009-0001-2462-2272}
\affiliation{%
  \institution{University of Illinois Urbana-Champaign}
 \city{Urbana}
 \state{IL}
 \country{USA}}
 \email{keranw2@illinois.edu}

\author{Violeta J. Rodriguez}
\orcid{0000-0001-8543-2061}
\affiliation{%
 \institution{University of Illinois Urbana-Champaign}
 \city{Champaign}
 \state{IL}
 \country{USA}}
 \email{vjrodrig@illinois.edu}

\author{Ravi Karkar}
\orcid{0000-0003-1467-4439}
\affiliation{%
 \institution{University of Massachusetts Amherst}
 \city{Amherst}
 \state{MA}
 \country{USA}}
 \email{rkarkar@umass.edu}

\author{Koustuv Saha}
\orcid{0000-0002-8872-2934}
\affiliation{%
 \institution{University of Illinois Urbana-Champaign}
 \city{Urbana}
 \state{IL}
 \country{USA}}
 \email{ksaha2@illinois.edu}

\renewcommand{\shortauthors}{}


\input{0abstract}


% \ccsdesc[300]{Human-centered computing~Social media}


\maketitle

% \ifhidecomments
% \else
%     \thispagestyle{firststyle} % applies firststyle to first page
%     \pagestyle{allstyle}
% \fi

% \input{0rough}
\input{1introduction.tex} %1.25 pages
\input{2rw_new.tex} %1 page
\input{3study} %2 page
\input{4.1results_interactions} %2 page
% \input{4.2results_needs} %2 page

\input{4.3results_alignment}
%2 
\input{5discussion_new} %2 page
\input{6limitations} %3 page
\input{7conclusion.tex}
%3 page


\begin{acks}
We sincerely thank all study participants for their time, insights, and willingness to share their experiences. This work was partly supported by the National Institute on Aging of the National Institutes of Health under Award Number P30AG073105. It was also partly supported by the Jump ARCHES endowment through the Health Care Engineering Systems Center at Illinois and the OSF Foundation.
\end{acks}

\bibliographystyle{ACM-Reference-Format}
\bibliography{mainArXiv}

% \includepdf[pages=-,pagecommand={},width=1.2\textwidth]{CSCW Job Satisfaction Reviewer Responses.pdf}

\appendix
\input{9appendix.tex}

\end{document}

\endinput
%%
%% End of file `sample-acmsmall.tex'.

%% file: 00packages.tex
\usepackage{longtable}
\usepackage{capt-of}
\usepackage{balance,bm}
\usepackage{color, colortbl, xcolor}
\definecolor{LightGray}{gray}{0.97}
\usepackage{pgfplots}
\usepackage{xcolor}
\usepackage{url}
\usepackage{makecell}

\usepackage{subcaption}
\usepackage{booktabs} % For formal tables
\usepackage{graphicx}
\usepackage{textcomp}
\usepackage{color,soul}
\usepackage{bm}
\usepackage{multirow}
\usepackage{wrapfig}
\usepackage{balance}
\usepackage{graphicx}  %Required
\usepackage{enumitem}

\usepackage{colortbl}
\usepackage{arydshln}
\usepackage [english]{babel}
\usepackage [autostyle, english = american]{csquotes}
\definecolor{linkColor}{RGB}{6,125,233}
\definecolor{green}{rgb}{0.0, 0.65, 0.31}
\definecolor{bleudefrance}{rgb}{0.19, 0.55, 0.91}
\definecolor{ceruleanblue}{rgb}{0.16, 0.32, 0.75}
\definecolor{grey}{HTML}{969696}
\definecolor{violet}{HTML}{756bb1}
\definecolor{dgrey}{HTML}{01665e}
\definecolor{lgrey}{HTML}{5ab4ac}
\definecolor{dgreen}{HTML}{005a32}
\definecolor{purple}{HTML}{ae017e}

\definecolor{editCol}{HTML}{000000}
\definecolor{maskCol}{HTML}{c51b7d}
\definecolor{lrColor}{HTML}{8856a7}
\definecolor{trColor}{HTML}{d01c8b}
\definecolor{ctColor}{HTML}{4dac26}
\definecolor{brickred}{HTML}{f03b20}
\definecolor{improveCol}{HTML}{253494}
\definecolor{worsenCol}{HTML}{d7191c}
\definecolor{DarkBlue}{HTML}{00008B}
\definecolor{mscolor}{HTML}{01665e}
\definecolor{nmscolor}{HTML}{bf812d}
\definecolor{lgreen}{HTML}{ccece6}
\definecolor{dolive}{HTML}{308014}

\definecolor{lrColor}{HTML}{8856a7}
\definecolor{trColor}{HTML}{d01c8b}
\definecolor{ctColor}{HTML}{4dac26}
\definecolor{brickred}{HTML}{f03b20}
\definecolor{improveCol}{HTML}{253494}
\definecolor{worsenCol}{HTML}{d7191c}
\definecolor{lgreen}{HTML}{e0f3db}
\definecolor{dpink}{HTML}{CD1076}
\definecolor{pink}{HTML}{FED2D2}
\definecolor{soothinggreen}{HTML}{4dac26}
\definecolor{darkred}{HTML}{8B0000}

\definecolor{dblue}{HTML}{104E8B}
\definecolor{violet}{HTML}{8A2BE2}
\definecolor{mscolor}{HTML}{01665e}
\definecolor{nmscolor}{HTML}{d8b365}
\definecolor{deepgrey}{HTML}{525252}
\definecolor{dslate}{HTML}{2F4F4F}
\definecolor{dolive}{HTML}{556B2F}
\definecolor{teal}{HTML}{388E8E}
\definecolor{mscolor}{HTML}{01665e}
\definecolor{nmscolor}{HTML}{d8b365}

\definecolor{aicolor}{HTML}{018571}
\definecolor{occolor}{HTML}{ff7799}

\definecolor{srcolor}{HTML}{e34a33}
\definecolor{smcolor}{HTML}{253494}
\definecolor{srsmcolor}{HTML}{7fcdbb}
\definecolor{bothcolor}{HTML}{fe9929}
\definecolor{onecolor}{HTML}{018571}
\definecolor{marroon}{HTML}{881c1c}

\usepackage{mathtools}

\usepackage{amsmath}
\usepackage{array}
\usepackage{xcolor}
\usepackage{arydshln}
\usepackage{siunitx}
\colorlet{tablerowcolor4}{gray!50} % Table row separator colour = 

\newcommand*{\textlabel}[2]{%
  \edef\@currentlabel{#1}% Set target label
  \phantomsection% Correct hyper reference link
  #1\label{#2}% Print and store label
}

\colorlet{tableheadcolor}{gray!25} % Table header colour = 25% gray
\colorlet{tablerowcolor}{gray!15} % Table row separator colour = 
\colorlet{tablerowcolor2}{gray!45} % Table row separator colour = 
\colorlet{tablerowcolor3}{gray!25} % Table row separator colour = 10% gray

\newcommand{\rowcollight}{\rowcolor{LightGray}} %
\usepackage{capt-of}
\usepackage{balance,bm}
\usepackage{color, colortbl, xcolor}

\usepackage{booktabs}  
\usepackage{url}
\usepackage{hyperref}
\hypersetup{
    colorlinks=true,
}
\usepackage{colortbl}
\usepackage{arydshln}
\usepackage [english]{babel}
\usepackage [autostyle, english = american]{csquotes}
\definecolor{linkColor}{RGB}{6,125,233}
\definecolor{green}{rgb}{0.0, 0.65, 0.31}
\definecolor{bleudefrance}{rgb}{0.19, 0.55, 0.91}
\definecolor{ceruleanblue}{rgb}{0.16, 0.32, 0.75}
\definecolor{grey}{HTML}{969696}
\definecolor{violet}{HTML}{756bb1}
\definecolor{dgrey}{HTML}{01665e}
\definecolor{lgrey}{HTML}{5ab4ac}
\definecolor{dgreen}{HTML}{005a32}
\definecolor{purple}{HTML}{ae017e}

\definecolor{editCol}{HTML}{000000}
\definecolor{maskCol}{HTML}{c51b7d}
\definecolor{lrColor}{HTML}{8856a7}
\definecolor{trColor}{HTML}{d01c8b}
\definecolor{ctColor}{HTML}{4dac26}
\definecolor{brickred}{HTML}{f03b20}
\definecolor{improveCol}{HTML}{253494}
\definecolor{worsenCol}{HTML}{d7191c}
\definecolor{DarkBlue}{HTML}{00008B}
\definecolor{mscolor}{HTML}{01665e}
\definecolor{nmscolor}{HTML}{bf812d}
\definecolor{lgreen}{HTML}{ccece6}
\definecolor{dolive}{HTML}{308014}

\definecolor{maskCol}{HTML}{c51b7d}
\definecolor{lrColor}{HTML}{8856a7}
\definecolor{trColor}{HTML}{d01c8b}
\definecolor{ctColor}{HTML}{4dac26}
\definecolor{brickred}{HTML}{f03b20}
\definecolor{improveCol}{HTML}{253494}
\definecolor{worsenCol}{HTML}{d7191c}
\definecolor{lgreen}{HTML}{e0f3db}
\definecolor{dpink}{HTML}{CD1076}
\definecolor{pink}{HTML}{FED2D2}
\definecolor{soothinggreen}{HTML}{4dac26}
\definecolor{darkred}{HTML}{8B0000}

\definecolor{dblue}{HTML}{104E8B}
\definecolor{violet}{HTML}{8A2BE2}
\definecolor{mscolor}{HTML}{01665e}
\definecolor{nmscolor}{HTML}{d8b365}
\definecolor{deepgrey}{HTML}{525252}
\definecolor{dslate}{HTML}{2F4F4F}
\definecolor{dolive}{HTML}{556B2F}
\definecolor{teal}{HTML}{388E8E}
\definecolor{mscolor}{HTML}{01665e}
\definecolor{nmscolor}{HTML}{d8b365}

\definecolor{aicolor}{HTML}{018571}
\definecolor{occolor}{HTML}{ff7799}

\definecolor{srcolor}{HTML}{e34a33}
\definecolor{smcolor}{HTML}{253494}
\definecolor{srsmcolor}{HTML}{7fcdbb}
\definecolor{bothcolor}{HTML}{fe9929}
\definecolor{onecolor}{HTML}{018571}
\definecolor{marroon}{HTML}{881c1c}

\usepackage{mathtools}

\usepackage{amsmath}
\usepackage{array}
\usepackage{xcolor}
\usepackage{arydshln}
\usepackage{siunitx}
\colorlet{tablerowcolor4}{gray!50} % Table row separator colour = 

\usepackage{tcolorbox}

  \def\nuLRbar#1{%%
  {\color{black}\rule{#1cm}{6pt}}
  }

\colorlet{tableheadcolor}{gray!25} % Table header colour = 25% gray
\colorlet{tablerowcolor}{gray!15} % Table row separator colour = 
\colorlet{tablerowcolor2}{gray!45} % Table row separator colour = 
\colorlet{tablerowcolor3}{gray!25} % Table row separator colour = 10% gray

\newif{\ifhidecomments}
  \hidecommentsfalse 
\ifhidecomments
    \newcommand{\keran}[1]{}
    \newcommand{\melissa}[1]{}
    \newcommand{\dongwhi}[1]{}
    \newcommand{\koustuv}[1]{}
    \newcommand{\ravi}[1]{}
\else
    \newcommand{\keran}[1]{\textbf{\small\sffamily{\textcolor{DarkBlue}{[#1 -- Keran]}}}}
    \newcommand{\melissa}[1]{\textbf{\small\sffamily{\textcolor{dolive}{[#1 -- Melissa]}}}}
    \newcommand{\ravi}[1]{\textbf{\small\sffamily{\textcolor{marroon}{[#1 -- Ravi]}}}}
    \newcommand{\dongwhi}[1]{\textbf{\small\sffamily{\textcolor{dpink}{[#1 -- Dong Whi]}}}}
    \newcommand{\koustuv}[1]{\textbf{\small\sffamily{\textcolor{violet}{[#1 -- Koustuv]}}}}
  \fi
\usepackage{xcolor}

\newcommand{\edit}[1]{#1}

\newcommand{\carey}{\textsf{Carey}}

\renewcommand{\textrightarrow}{$\rightarrow$}

\newcommand{\n}[1]{$\mathtt{#1}$}

\newcommand{\tabitem}{\textbullet~~}

\colorlet{tableheadcolor}{gray!25} % Table header colour = 25% gray
\colorlet{tablerowcolor}{gray!5} % Table row separator colour = 10% gray

\definecolor{neutralCol}{HTML}{dd1c77}
\definecolor{neutralGreen}{HTML}{31a354}
\definecolor{NewBlue}{HTML}{1879ba}
\definecolor{bleudefrance}{rgb}{0.19, 0.55, 0.91}  
\definecolor{AfTrColor}{HTML}{0868ac}  
\definecolor{BfTrColor}{HTML}{a8ddb5}  

\definecolor{AfCtColor}{HTML}{b10026}  
\definecolor{BfCtColor}{HTML}{fd8d3c}

\graphicspath{ {figures/} }

\newcommand{\para}[1]{\vspace{0.3em}\noindent\textbf{#1}~}

%% file: 0abstract.tex
\begin{abstract}
% \koustuv{Abstract needs to be modified}
% Alzheimer's Disease and Related Dementias (AD/ADRD) are progressive neurodegenerative conditions that impair memory, thought processes, and functioning. 
% Family caregivers of individuals with AD/ADRD face significant mental health challenges owing to their long-term caregiving responsibilities. 
% Yet current support systems often overlook the evolving nature of their mental wellbeing needs. 
% Our study examines the caregivers' mental wellbeing concerns, focusing on the practices they adopt to manage the caregiving burden and the technologies they use for support. Through semi-structured interviews with 25 family caregivers of individuals with AD/ADRD, we identified the key causes and effects of mental health challenges and uncovered the evolution of caregivers' mental wellbeing over time. 
% Additionally, our participants shared insights into improvements for existing mental health technologies, emphasizing the need for accessible, scalable, and personalized solutions. 
% These findings offer a foundation for designing tailored interventions that holistically support caregivers' mental wellbeing, benefiting both caregivers and care recipients. 
Family caregivers of individuals with Alzheimer's Disease and Related Dementia (AD/ADRD) face significant emotional and logistical challenges that place them at heightened risk for stress, anxiety, and depression. Although recent advances in generative AI---particularly large language models (LLMs)---offer new opportunities to support mental health, little is known about how caregivers perceive and engage with such technologies. 
To address this gap, we developed \carey{}, a GPT-4o–based chatbot designed to provide informational and emotional support to AD/ADRD caregivers. 
Using \carey{} as a technology probe, we conducted semi-structured interviews with 16 family caregivers following scenario-driven interactions grounded in common caregiving stressors. 
Through inductive coding and reflexive thematic analysis, we surface a systemic understanding of caregiver needs and expectations across six themes---\textit{on-demand information access, safe space for disclosure, emotional support, crisis management, personalization}, and \textit{data privacy}. 
For each of these themes, we also identified the nuanced tensions in the caregivers' desires and concerns. 
We present a mapping of caregiver needs, AI chatbots' strengths, gaps, and design recommendations. 
Our findings offer theoretical and practical insights to inform the design of proactive, trustworthy, and caregiver-centered AI systems that better support the evolving mental health needs of AD/ADRD caregivers.

% health challenges but often lack access to affordable, personalized, and timely support. This paper explores the potential role of generative AI in supporting caregiver wellbeing through \carey{}, a large language model (LLM)-powered chatbot designed to provide emotional and practical support. 
% \carey{} is designed to augment existing caregiving and mental health resources--not replace professional or clinical care--by offering accessible, emotionally attuned support for caregivers. 
% Using a scenario-based interview study with 16 AD/ADRD caregivers, we examine how caregivers perceive \carey{}'s usefulness, trustworthiness, and emotional alignment. Through reflexive thematic analysis of interview transcripts and chatbot interactions, we identify key alignments—such as caregivers valuing nonjudgmental validation, 24/7 availability, and low-barrier emotional expression—as well as critical misalignments around information accuracy, lack of initiative, and limited personalization. Our findings offer design recommendations for human-centered AI systems that support the evolving mental health needs of caregivers and contribute to broader conversations on trust, empathy, and ethics in AI-mediated wellbeing technologies.
\end{abstract}

%% file: 1introduction.tex
\section{Introduction}\label{section:intro}
% \koustuv{@melissa, could you have a pass of the introduction.}

Alzheimer's Disease (AD) and Alzheimer's Disease-Related Dementias (ADRD) are one of the most pressing public health challenges---as of 2024, 6.9 million people in the U.S. aged 65 and older \edit{are} living with AD/ADRD\footnote{We use the term AD/ADRD to be consistent with terminology used by the National Alzheimer's Project Act~\cite{adrd2025napa}.}, a number projected to nearly double to 13.8 million by 2050 as the population ages~\cite{nichols2022estimation,alzheimer2024facts}.
The progressive nature of these neurodegenerative conditions leads to cognitive decline, memory loss, and increasing dependence on others for daily functioning~\cite{better2023alzheimer}. 
The majority of care is provided by informal caregivers~\cite{garcia2011anxiety,grunfeld1997caring}---primarily family members---who, in 2023, delivered over 18.4 billion hours, or nearly 31 hours per week, of unpaid care~\cite{alzheimer2024facts}. 
Family caregivers are essential to maintain the wellbeing and safety of individuals living with dementia, but they often do so under conditions of high emotional strain, minimal respite, and limited access to professional resources~\cite{unbehaun2018exploring,gibson2014exploring,shi2025balancing}. 
Compared to formal caregivers, family caregivers face more pronounced challenges due to factors such as emotional attachment to the care recipient, lack of formal training, and insufficient resources and equipment~\cite{duplantier2023barriers,reinhard2008supporting}.
As a result, they often struggle with high levels of stress, anxiety, and depression
% ~\cite{pinquart2003associations, livingston1996depression,hughes2014correlates,cho2023effect}
---at least one in three family caregivers of AD/ADRD individuals suffer from clinical depression per prior meta-analyses~\cite{sallim2015prevalence}.
% ,ying2019validity}. 
% While depression and anxiety are common to caregivers, a
Additional challenges such as anticipatory grief, compassion fatigue, caregiver guilt, and role-related identity shifts~\cite{brodaty2009family} uniquely characterize caregivers'\footnote{Since our work focuses on family caregivers, the term ``caregivers'' will refer to family caregivers unless otherwise noted.} mental health experience.

Despite the growing recognition of these mental health challenges, access to timely, affordable, and personalized support remains limited for most caregivers~\cite{caputo2021emotional}. 
% ~\cite{novais2017explore, caputo2021emotional}. 
Traditional mental health services can be costly, stigmatizing, or difficult to access due to time constraints and caregiving responsibilities~\cite{clemmensen2021know}. 
In response, researchers and designers have begun exploring the role of technology in enhancing mental health support~\cite{fitzpatrick2017delivering, seitz2024artificial}. 
Among these, conversational agents and AI-driven chatbots present a particularly promising avenue due to their accessibility, adaptability, and ability to simulate supportive dialogue~\cite{zhu2023toward, miner2016smartphone}.
\edit{Rather than replacing human relationships, such tools are framed as low-burden, on-demand complements that can support caregivers when other forms of support are unavailable, difficult to access, or emotionally taxing to seek.}

Recent advances in generative artificial intelligence (AI), especially large language models (LLMs), offer an opportunity to enhance caregiving support~\cite{topol2019high}. 
LLM-powered chatbots can provide always-available support, practical advice, and a sense of companionship~\cite{inkster2018empathy, zheng2023building, chen2023soulchat}. 
However, little is known about how AD/ADRD caregivers perceive such technologies, and to what extent they trust, use, or feel supported by them---especially in the context of mental health and wellbeing. 
Prior work has highlighted how user expectations of technology evolve alongside their needs and contexts~\cite{cai2019hello}. 
Therefore, it is essential to examine not only whether AI can support caregivers, but also \textit{how caregivers interpret, negotiate, and assess the role of AI in their mental wellbeing journey}. 
In fact, these needs and expectations are not merely preferences---they are foundational to designing human-centered AI systems that meaningfully support caregiver mental health.

To bridge the above gap, we investigate the research question (RQ): \textbf{What are the needs and expectations of AD/ADRD caregivers for an AI chatbot to support mental health?}

For our RQ, we developed a GPT-4o based AI chatbot, named \carey{}---which is aims at providing informational and emotional support to AD/ADRD caregivers. 
We used \carey{} as a technology probe~\cite{hutchinson2003technology} to conduct semi-structured interviews with 16 AD/ADRD caregivers.
% an AI chatbot-based technology probe~\cite{hutchinson2003technology}, named \carey{}.
% and conducted semi-structured interviews with 16 AD/ADRD caregivers. 
% \carey{} is designed to support existing caregiving and mental health resources---not replace professional or clinical care---by offering accessible, emotionally attuned support for caregivers. 
% We conducted semi-structured interviews with 16 AD/ADRD caregivers who 
% Participants 
These participants interacted with \carey{} on eight carefully designed scenarios from the literature, representing common caregiving stressors---1) disruptive care-recipient's behavior, 2) lack of support, 3) low self-efficacy, 4) emotional distress, 5) relationship tensions, 6) compassion fatigue, 7) lack of self-care, and 8) burnout.
The participants shared their interaction experiences and perceptions about their expectations, desires, and concerns about \carey{}.
\edit{Consistent with a scenario-driven technology-probe approach, our goal was not to evaluate long-term use or real-world effectiveness, but to elicit caregivers' perceptions, expectations, and concerns through guided interactions.}

% such as emotional distress, compassion fatigue, or lack of support---and shared their perceptions of \carey{}'s effectiveness, trustworthiness, and emotional resonance. 
% Accordingly, our research investigates the expectations from an AI of AD/ADRD caregivers\footnote{Since our work focuses on family caregivers, the term 'caregivers' will refer to family caregivers unless otherwise noted.}. Our study is guided by the following research questions (RQs):
% \koustuv{add another sentence about the details of our findings, e.g., we present a systemic understanding of..} \melissa{done} 

We analyzed the interview transcripts and chat history using inductive coding and reflexive thematic analysis~\cite{braun2019reflecting} to examine both benefits and tensions in caregivers' interactions with \carey{}. 
We present a systemic understanding of the multifaceted ways caregivers seek, interpret, and evaluate AI-based support in response to specific emotional and logistical stressors. 
Our findings highlight both optimism and hesitation---while caregivers appreciated \carey{}'s timeliness, ease of access, and non-judgmental tone, they also raised concerns about personalization, initiative, information trust, and emotional over-reliance. 
Importantly, caregivers expressed that their mental wellbeing support needs are contextual, evolving, and relational---dependent not only on their own emotional states, but also on interactions with care recipients, family dynamics, and external resources. 
\edit{Taken together, our findings suggest that tools like \carey{} can be envisioned as low-burden, on-demand supports that caregivers may turn to during brief or constrained moments---such as late at night, between caregiving tasks, or when other forms of support are unavailable---rather than as sustained or time-intensive interventions.}
Our work offers the following key contributions:

\begin{itemize}
\item We provide a systemic understanding of how caregivers engage with an AI mental health chatbot, surfacing key needs and expectations across six major themes---\textit{on-demand information access, emotional support, safe space for disclosure, crisis management, personalization capabilities}, and \textit{data privacy}. 

\item For each of the above themes, we highlight the caregivers' nuanced perceptions of benefits and concerns.

\item We present a comprehensive mapping of caregiver needs, current chatbot strengths, unmet gaps, and actionable design recommendations---centered around the caregiver experiences.

% \item We discuss the implications of our work in informing the development of caregiver-centered AI tools that are proactive, trustworthy, and attuned to the emotional and informational complexities of caregiving.
\end{itemize}

% \begin{itemize}
% \item We provide a systemic understanding of how caregivers engage with an AI mental health chatbot, highlighting their needs and expectations, across six major themes of on-demand information access, safe-space for disclosure, emotional support, crisis management, and personalization capabilities.
% \item We provide a mapping of needs, strengths, gaps, and design recommendations to guide the development of caregiver-centered AI tools for mental health.
% % \item We offer insights to inform the design of human-centered AI tools that are emotionally responsive, context-aware, and trustworthy.
% \end{itemize}

% \begin{itemize}
% \item \edit{We provide a systemic understanding of how caregivers engage with an AI mental health chatbot \carey{}, highlighting needs for emotional support, trustworthy information, and conversational naturalness.} 
% \item \edit{We map the key alignments and misalignments between \carey{} and caregiver expectations, revealing design tensions in trust, initiative, and ethical guidance.}
% \item \edit{We offer design recommendations for future caregiver-focused AI systems that center on adaptability, proactive empathy, contextual awareness, and data privacy.} 
% \end{itemize}

The above contributions collectively advance both theoretical understanding and practical approaches to supporting AD/ADRD caregivers' mental wellbeing. 
We discuss the implications of our work in informing the development of caregiver-centered AI tools that are proactive, trustworthy, and attuned to the emotional and informational complexities of caregiving.
% By illuminating how caregivers interpret and emotionally respond to AI-provided support, our findings inform the design of more effective, trustworthy, and emotionally attuned AI systems in caregiving and mental health domains. 
Our work underscores the importance of moving beyond technical feasibility toward context-sensitive, relationally grounded AI support for vulnerable populations.

%% file: 2rw_new.tex
\section{Background and Related Work}\label{section:rw}
\subsection{Alzheimer's Disease and Related Dementias (AD/ADRD): Condition and Caregiving}
Alzheimer's Disease and Related Dementias (AD/ADRD) is a group of progressive neurodegenerative conditions marked by cognitive decline, memory loss, and eventual loss of independence~\cite{kim2021scoping}. 
With no known cure, AD/ADRD remains a significant and growing public health challenge---currently the fifth-leading cause of death among older adults in the U.S.~\cite{better2023alzheimer}. By 2050, it is estimated that 14 million people in the U.S. and 152 million globally will be living with these conditions~\cite{nichols2022estimation}.

Care for individuals with AD/ADRD is primarily provided by family and informal caregivers, most often within the home environment~\cite{garcia2011anxiety,grunfeld1997caring}. 
In 2022, this unpaid caregiving was valued at over \$339.5 billion in the U.S. alone~\cite{better2023alzheimer}. 
This caregiving experience is marked by significant emotional, physical, and financial strain. Caregivers frequently report challenges in managing their own mental and physical health~\cite{coon2009empirically,hughes2014correlates,wawrziczny2017spouse}, navigating future care decisions~\cite{millenaar2018exploring,vaingankar2013perceived}, and maintaining financial stability~\cite{gibson2014exploring,griffiths2016problems,unbehaun2018exploring}. The intensity of AD/ADRD caregiving, including the need to supervise daily activities, manage behavioral symptoms, and make complex medical decisions, contributes to elevated stress levels and heightened risk of depression and burnout~\cite{harper2022,alzheimer2005alzheimer,rettinger2020mixed,queluz2020understanding,lee2022unmet}.

% Compared to caregivers who support individuals with other health conditions, AD/ADRD caregivers often face more pronounced and prolonged emotional demands~\cite{alzheimer2005alzheimer}. 
The daily routines of AD/ADRD caregivers are often tightly constrained by the needs of the care recipient, leaving limited time for self-care, social engagement, or accessing professional support~\cite{rettinger2020mixed, wawrziczny2016needing}. 
This localization of caregivers' lives---often physically, emotionally, and socially---further compounds their vulnerability, with many reporting isolation, limited mobility, and difficulty accessing even basic mental health resources~\cite{harper2022,alzheimer2005alzheimer,rettinger2020mixed}.
They are at a heightened risk for clinical concerns such as suicidal ideation, complicated grief, trauma symptoms, and health-related anxieties~\cite{schulz2016families, sorensen2011issues}. 
% These elevated risks reinforce the urgency of providing accessible, contextually sensitive emotional support. 
% \koustuv{We should add how their lives are often limited/localized around the care-recipient which leads to a lack of sufficient time to seek active care, lack of social lives, etc.}
% Further, their daily routines become tightly constrained around the care recipient’s needs, leaving little time for self-care, social engagement, or seeking professional support~\cite{rettinger2020mixed, wawrziczny2016needing}. 
Without formal training or structured support, they must adapt to the evolving needs of the care recipient while contending with their own wellbeing challenges. 

% \hl{We examine not only the causes and consequences of psychological distress but also how these experiences evolve across the caregiving journey. By highlighting caregivers' dynamic needs and the barriers they face in managing emotional health, our study contributes a more nuanced understanding of how support systems and technologies can be designed to address the complex realities of dementia caregiving.} 
% \koustuv{these two sentences were more suited for the CSCW submission, for this one we need to talk more about an AI chatbot-based intervention.}

A growing body of research acknowledges that caregiving is not just a logistical challenge, but also an emotionally complex and socially entangled experience~\cite{harper2022,garcia2011anxiety,grunfeld1997caring,better2023alzheimer}.
Caregiving in AD/ADRD contexts is particularly complex, involving prolonged emotional labor, progressive memory loss, behavioral challenges, and shifting family roles~\cite{brodaty2009family, chen2022experiences}.
This complexity is well-documented in clinical frameworks such as Pearlin's Stress Process Model~\cite{pearlin1981stress}, which outlines how contextual stressors, mediators (e.g., social support), and outcomes (e.g., depression, burden) dynamically interact throughout the caregiving trajectory.
Given these challenges, there is an urgent need for interventions that are accessible, affordable, immediate, and available around-the-clock---especially given caregivers' limited time and mobility, and the unpredictability of care demands. 
AI-driven tools, such as conversational agents, offer a promising direction for delivering low-barrier scalable support~\cite{xie2020artificial,wang2024artificial}. 
However, there is limited understanding of what AI-driven mental wellbeing support should look like for AD/ADRD caregivers.
Building on the above body of research, our work focuses specifically on the mental wellbeing of family caregivers in AD/ADRD contexts.  
% Drawing on the above body of literature, w
We designed scenario-based interactions to surface caregivers' perspectives, exploring their expectations, concerns, and perceived value of AI-driven support through a chatbot interface.

\subsection{HCI and Digital Health Technologies for Caregivers}
Human-Computer Interaction (HCI) and Digital Health research has long examined how digital technologies can support the wellbeing of caregivers~\cite{bosch2019caregiver, miller2016partners, chen2013caring, seo2019balancing, jacobs2019think, lee2023reimagining, schorch2016designing, zubatiy2021empowering,fu2025felt}. 
In the context of caregiving for AD/ADRD and older adults, prior work has examined the use of smartphone~\cite {shreve2016dementia}, wearables~\cite{kourtis2019digital,stavropoulos2021wearable}, voice assistants~\cite{wong2024voice,o2020voice,pradhan2020use}, and online social platforms~\cite{johnson2022s,levonian2021patterns,pickett2024carevirtue,kaliappan2025online,saha2025ai} to reduce caregiver burden, facilitate coordination, and enhance access to resources. 
Prior research has explored technologies such as memory aids, context-aware home systems~\cite{van2017assistive}, and collaborative care platforms~\cite{johnson2022s,levonian2021patterns} to assist both caregivers and care recipients. 
These technologies often aim to support caregiving tasks such as medication management~\cite{lee2022unmet}, appointment scheduling~\cite{boessen2017online}, and activity monitoring, while also supporting emotional needs through self-tracking, social support, or reflective prompts~\cite{boessen2017online,kaliappan2025online}. 

Beyond functional support, a rich body of work has emphasized the importance of addressing the emotional and psychological dimensions of caregiving~\cite{siddiqui2023exploring,lazar2017supporting,bhat2023we,shi2025balancing}. \citeauthor{lazar2017supporting} demonstrated the value of fostering digital social sharing among AD/ADRD caregivers to combat isolation and enhance communal coping~\cite{lazar2017supporting,johnson2022s}. \citeauthor{bhat2023we} highlighted the caregiver's central role in mediating care and proposed caregiver-centric technological supports~\cite{bhat2023we}. Similarly,~\citeauthor{kim2024opportunities} pointed to the need for adaptive and personalized support strategies, given that caregiver needs often fluctuate across the progression of behavioral episodes~\cite{kim2024opportunities}. \citeauthor{meyerhoff2022meeting} advocated for user-centered digital mental health tools that flexibly adapt to individual support needs and empower users in their mental health journeys~\cite{meyerhoff2022meeting}. Most recently, \citeauthor{smriti2024emotion} explored how technology might meaningfully address the emotional experiences of dementia caregivers~\cite{smriti2024emotion}.

\edit{Accordingly, recent research has called for digital interventions that center caregivers' wellbeing, relational dynamics, and long-term coping strategies~\cite{yoo2024patient,bhat2023we,shi2025balancing}.
Mental health chatbots, journaling platforms, and peer support communities have shown early promise in reducing emotional burden. However, how caregivers perceive, trust, and engage with conversational AI---particularly in relation to their evolving mental health needs---remains underexplored.}
Our work builds on this evolving literature by examining how an AI-powered mental health chatbot may fulfill---or fall short of---caregivers' expectations, with the goal of identifying actionable design considerations for human-centered technologies to support caregivers' mental wellbeing.

\subsection{Designing Human-Centered AI for Wellbeing: Maximizing Benefits and Mitigating Harms}

AI holds growing promise for addressing longstanding gaps in mental health and caregiving support~\cite{borna2024artificial}.
Increasingly, AI systems are being leveraged to deliver mental and physical health support across diverse domains, including clinical decision support, health behavior change, symptom tracking, and emotional wellbeing~\cite{luxton2015artificial, topol2019high,slovak2023designing,jo2023understanding}.
\edit{Prior research has explored a diverse range of AI-powered modalities for mental health and wellbeing support, including sensing-based systems, recommender systems, and generative media tools~\cite{yoo2024patient,slade2024implementation, adler2024beyond}. For example, generative models beyond text---such as text-to-image systems---have been explored to facilitate emotional expression and self-disclosure in online mental health contexts~\cite{zhang2025mentalimager,steijger2025use}.}

Among these, AI chatbots have emerged as accessible, low-cost tools that aim to provide emotional support at scale~\cite{inkster2018empathy,ibrahim2025designing}.
These tools can integrate techniques such as Cognitive Behavioral Therapy (CBT)~\cite{beck1976cognitive,sharma2024facilitating} and emotion regulation~\cite{slovak2023designing}, and have shown early promise in reducing symptoms of depression and anxiety~\cite{fitzpatrick2017delivering}.

With the advancements in large language models (LLMs), interest in AI-driven mental health technologies has expanded significantly~\cite{xu2024mental}.
General-purpose LLM-based chatbots such as OpenAI's ChatGPT and Google's Gemini are now being explored as companions or wellbeing coaches due to their ability to generate emotionally expressive and contextually adaptive responses~\cite{zheng2023building, chen2023soulchat}.
This technological shift has intensified hopes that generative AI could play a central role in expanding mental health support, particularly for caregivers whose lives are constrained by the continuous demands of care.

\edit{In parallel, recent reviews highlight the breadth of AI applications in dementia-related contexts, including tools designed to support quality of life, care coordination, and wellbeing for both people living with dementia and their caregivers~\cite{steijger2025use}. However, much of this work has focused on task support or patient-facing interventions, with comparatively less attention to how caregivers engage with conversational AI systems for their own mental health needs.}
\edit{LLM-based conversational agents are especially relevant in caregiver contexts because they support open-ended dialogue, allow caregivers to articulate complex emotional states in natural language, and provide on-demand interaction without requiring sustained attention or structured inputs. These properties distinguish LLMs from other AI modalities and motivate our focus on conversational systems in this study.}

Yet, despite these advances, significant concerns remain~\cite{de2023benefits,cabrera2023ethical,yoo2025ai,saha2025ai,goel2026rubrix}. LLMs are known to hallucinate facts, produce inconsistent or overly generic responses, and simulate empathy without true understanding~\cite{bender2021dangers}.
They struggle to detect psychological distress, adapt to evolving user needs, or maintain memory over time---capabilities that are crucial in emotionally sensitive, high-stakes domains like dementia caregiving.
These limitations risk undermining user trust and may inadvertently cause harm.
Further concerns around bias, safety, and ethical accountability highlight the need for careful, intentional design~\cite{liu2023trustworthy}.

As AI systems move beyond clinical settings into everyday wellbeing contexts, HCI and digital mental health researchers have emphasized the importance of human-centered design principles~\cite{chancellor2019taxonomy,stade2024large,schroeder2019examining}.
Prior work has called for tailoring AI to individual needs—not just to optimize outcomes, but also to support emotional processes such as expression, validation, and coping~\cite{cai2019hello, berretta2023defining}.
% \citeauthor{madaio2020co} proposed participatory co-design of ethical AI checklists, developed with stakeholders to address organizational challenges and surface hidden biases~\cite{madaio2020co}.
Prior research has called for greater accountability, inclusive, and value-driven AI design~\cite{ehsan2023charting,kawakami2023wellbeing,yoo2024missed,das2024teacher}.
Likewise, research on AI alignment---ensuring that AI systems produce outcomes consistent with human values---has gained increasing attention, expanding from its roots in philosophy and machine learning to more human-centered design~\cite{gabriel2020artificial, terry2023ai}.
% Recent work highlights alignment both during development (e.g., annotation) and interaction (e.g., user feedback)~\cite{li2023coannotating, chung2022talebrush,ma2023should}, with concepts like bidirectional alignment emphasizing mutual adaptation between humans and AI~\cite{shen2024towards}.

That said, in the domain of AD/ADRD caregiving, chatbot technology remains in its early stages and is not yet well aligned with caregivers’ lived experiences, emotional labor, or evolving support needs~\cite{ruggiano2021chatbots, wong2024voice}.
To fulfill the promise of AI for caregiver wellbeing, systems must be designed with deep consideration for trust, usability, emotional resonance, and harm reduction.
Our study builds on this body of work by investigating how AD/ADRD caregivers engage with a prototype LLM-powered mental health chatbot.
We provide a mapping of caregivers' needs, the chatbot's strengths and gaps, and resulting design considerations.
This mapping can inform the development of AI systems that are not only effective but also ethically aligned with the lived experiences and needs of AD/ADRD caregivers.

%% file: 3study.tex
\section{Study Design and Methods}
Our study used a scenario-based interview study with a technology probe~\cite{hutchinson2003technology}---\carey{}, a chatbot designed for AD/ADRD caregivers. This study was approved by the Institutional Review Boards (IRBs) at the researchers' institutions. In the following subsections, we describe our methodology.

\subsection{Participant Recruitment}

% \subsubsection{Recruitment}
To access a diverse and digitally engaged caregiver population, we recruited participants from social media communities, where many AD/ADRD caregivers seek support and share lived experiences~\cite{kaliappan2025online}.
We first contacted the moderators of the online communities catering to AD/ADRD-related discussions on Reddit (\textit{r/alzheimers, r/dementia, r/dementiaresearch, r/ParentsWithAlzheimers}, etc.), \textit{alzheimersdisease.net}, and \textit{alzconnected.org}, by briefly describing our research and if they were okay with recruiting from their respective platforms. 
In communities where the moderators granted permission, we posted recruitment information with an interest form. 
This form included a demographic questionnaire (collecting information on age, sex, race, U.S. state) and a section identifying their caregiving role. 
The form served as a screening tool to ensure participants met the eligibility criteria: 1) 18 years or older, 2) current or former caregivers for individuals with AD/ADRD, and 3) residing in the U.S. 
We received 205 responses to our interest form over three months from October 2024 to January 2025, and we invited a subset of participants to maximize diversity and balance across demographic attributes. 
% Ultimately, 16 participants consented and were interviewed. 
Ultimately, we interviewed 16 participants, each of whom was compensated with Amazon gift vouchers of \$25 USD.~\autoref{tab:participants} summarizes our participant pool, reflecting a balanced distribution across demographic attributes. Our sample includes both recent caregivers with less than three years of experience (\n{N}=9) and more experienced caregivers with over three years of caregiving experience (\n{N}=7).

\begin{table}[t]
\centering
\sffamily
\footnotesize
   \caption{Summary of participants, including type (current/former caregiver), years of caregiving (ys.), care-recipient, age, gender, race, education, and technology used for mental wellbeing (if any). \textit{Professional Caregivers} are marked with an `*' next to their ID.}
   \label{tab:participants}
\setlength{\tabcolsep}{1pt}
\resizebox{\columnwidth}{!}{
\begin{tabular}{llrlrllllll}
\textbf{ID} & \textbf{Type} & \textbf{Ys.} & \textbf{Care Recipient} & \textbf{Age} & \textbf{Sex} & \textbf{Race} & \textbf{Education} & \textbf{Tech. Used} \\
\toprule
P1  & Current & 4  & Father     & 36-50  & Transgender & Black or African American & Some college, no degree & 
\edit{Computer, smartphone apps, social media}\\
\rowcollight P2  & Current & 5  & Father     & 25-35  & Male        & Black or African American & Associate degree         &  \edit{Smartphone, laptop, ChatGPT, Headspace}\\
P3  & Current & 3  & Father     & 50-65  & Female      & White                      & Advanced degree         &  \edit{Smartphone apps} \\
\rowcollight P4  & Former  & 4  & Aunt       & 25-35  & Female      & Black or African American & Some college, no degree &  \edit{Social media} \\
P5  & Current & 5  & Stepmother & 25-35  & Male        & White                      & Advanced degree         & \edit{Computer} \\
\rowcollight P6  & Current & 3  & Father     & 25-35  & Male        & Hispanic or Latino         & Some college, no degree & \edit{YouTube, podcast} \\
P7  & Current & 7  & Mother     & 25-35  & Female      & Black or African American & Bachelor's degree       &  \edit{Smartphone, meditation apps, social media, online counseling} \\
\rowcollight P8  & Former  & 5  & Uncle      & 36-50  & Female      & White                      & Advanced degree         &   \edit{Social media, smartphone} \\
P9  & Current & 1  & Mother     & 50-65  & Female      & White                      & Bachelor's degree       &  \edit{Smartphone} \\
\rowcollight P10 & Current & 1  & Wife       & 50-65  & Male        & White                      & Advanced degree         & \edit{---} \\
P11 & Former  & 3  & Father     & 50-65  & Male        & Black or African American & Bachelor's degree       & \edit{Social media, mobile apps, smartphone} \\
\rowcollight P12 & Current & 3  & Father     & 19-24  & Female      & White                      & Some high school, no diploma &  \edit{Social media} \\
P13 & Current & 3  & Mother     & 50-65  & Female      & White                      & Associate degree        & \edit{Social media, alzheimers.org} \\
\rowcollight P14 & Current & 2  & Father     & 25-35  & Male        & Black or African American & Technical/Vocational training &  \edit{Online community, Headspace, Calm, YouTube, Instagram} \\
P15* & Current & 19 & Friend, Patients & 50-65 & Female   & White                      & Bachelor's degree       &  \edit{Smartphone, computer, Touchpoints} \\
\rowcollight P16 & Current & 3  & Mother    & 19-24  & Female      & Prefer not to say          & Associate degree        & \edit{Social media} \\
\end{tabular}}
\Description[table]{}
\end{table}

Before the interviews, participants were provided with the Rapid Caregiver Well-being Scale (R-CWBS)~\cite{tebb2013caregiver}. 
% along with the consent form. 
R-CWBS is a validated short-form rapid assessment instrument to infer key areas of support a caregiver needs~\cite{tebb2013caregiver}. 
Here each question is rated on a Likert-scale between 1 (Rarely) and 5 (Usually), and lower scores indicate a need for greater support. \autoref{table:cwbs} provides a summary of participants' responses to this survey---we see that our pool of participants scored high on taking care of personal daily activities (mean=4.25), but scored low on treating or rewarding themselves (mean=2.75).

Participants were also asked to respond to a questionnaire on the different mental health concerns of AD/ADRD caregivers drawn from the literature (the eight scenarios) on a Likert scale of 1 (not at all concerning) to 5 (very concerning).
~\autoref{table:mentalwellbeing_scale} summarizes their responses---we see that lack of support (mean=3.89) and burnout (mean=3.75) scored the highest, whereas low self-efficacy (mean=2.94) scored the least among our participants.

% \begin{table}[t!]
% \centering
% \sffamily
% \footnotesize
% % \caption{Mean Responses for Each Question (Numerical)}
% \caption{Summary of participants' responses to prompts on mental wellbeing concerns from literature. They responded to these prompts based on their level of concern, on a scale of 1 (not at all concerning) to 5 (very concerning).}
% \label{table:mentalwellbeing_scale}
% \begin{tabular}{lrrc}
% \textbf{Question} & \textbf{Mean} & \textbf{Std. Dev.} & \textbf{Distribution}\\
% \toprule
% % \rowcollight \multicolumn{4}{c}{\textbf{Activities}}\\
% Disruptive behaviors by care-recipient & 3.19 & 1.42 & \includegraphics[height=8pt]{figures/spike_mw_Disruptive.pdf}\\
% \rowcollight Lack of support & 3.89 & 1.20 & \includegraphics[height=6pt]{figures/spike_mw_Support.pdf}\\
% Low self-efficacy & 2.94 & 1.29 & \includegraphics[height=6pt]{figures/spike_mw_Self-Efficacy.pdf}\\
% \rowcollight Emotional distress & 3.44 & 1.42 & \includegraphics[height=6pt]{figures/spike_mw_Emotional Wellness.pdf}\\
% Relationship tensions & 3.38 & 1.20 & \includegraphics[height=6pt]{figures/spike_mw_Relationship.pdf}\\
% \rowcollight Compassion fatigue & 3.56 & 1.31 & \includegraphics[height=6pt]{figures/spike_mw_Compassion.pdf}\\
% Lack of self-care & 3.56 & 1.46 & \includegraphics[height=6pt]{figures/spike_mw_Self-Care.pdf}\\
% \rowcollight Burnout & 3.75 & 1.00 & \includegraphics[height=6pt]{figures/spike_mw_Burnout.pdf}\\
% % \bottomrule
% \end{tabular}
% \end{table}

\subsection{\carey{}: Technology Probe for AD/ADRD Caregivers' Mental Health}
Understanding the mental health challenges faced by AD/ADRD caregivers requires an approach that captures their caregiving experiences while interacting with AI-driven mental health tools. 
To explore this, we developed \carey{}---a GPT-4o-based chatbot---as a \textit{technology probe}~\cite{hutchinson2003technology}, \edit{a research approach commonly used in HCI to elicit users’ early reactions to an emerging or envisioned technology. Technology probes are not intended to evaluate long-term effectiveness or real-world adoption, but rather to support formative inquiry and generate design insights grounded in initial use.}
% a technology probe named \carey{}---a GPT-4o-mini-based chatbot---
\carey{} was designed to provide emotional support, practical caregiving advice, and self-care recommendations, \edit{and served as a research instrument to gather early insights into how caregivers perceive and engage with AI-based mental health support.}
% \carey{} serves as both an intervention and a research tool, allowing us to examine how caregivers perceive and engage with AI-based mental health support, and to explore how future technologies can better support them.
% \edit{We adopted a technology probe approach~\cite{hutchinson2003technology}, a method widely used in HCI to elicit participants' early reactions, expectations, and mental models around emerging technologies. Technology probes are particularly suited for formative design research, where the goal is not to evaluate long-term effectiveness or behavioral outcomes, but to surface values, tensions, and design opportunities that can inform future system development. In this study, \carey{} functioned as a probe to support reflection and discussion, enabling caregivers to articulate how such an AI system could fit into---or conflict with---their everyday caregiving practices and mental health needs.}
At the same time, due to the sensitive and complex nature of AD/ADRD caregiving and mental health, we opted not to deploy \carey{} in a real-world setting. 
Instead, participants interacted with Carey during a guided one-hour interview session. 
During the interviews, each participant was presented with eight caregiving-related scenarios---derived from the literature---which are known to significantly impact AD/ADRD caregivers' mental health. 
They were asked to select four to five of these scenarios as time permitted. 
% derived from previous literature and asked to select the four most relevant to their experiences. 
By engaging with \carey{} within these scenarios, participants shared their perspectives in a controlled setting, allowing for in-depth exploration of their needs and concerns. 
In the following subsections, we elaborate on the design of the scenarios and chatbot interactions.

% The following subsections elaborate on the scenario design process and the chatbot's development.

\subsubsection{Scenario Design}
We adopted a scenario-based design to guide our participants in interacting with \carey{} for mental health support. 
% These scenarios were carefully developed to avoid asking participants to directly share their personal caregiving experiences, which could be sensitive, unpredictable, and emotionally triggering. 
% Instead, we presented hypothetical but 
These theory-driven scenarios provided realistic caregiving situations that reflect common mental health stressors. 
For ethical reasons, interviewers monitored emotional cues and offered breaks or termination options when signs of distress emerged. 
Participants were also reminded that \carey{} is not a crisis resource. 
We developed the scenarios based on prior literature on AD/ADRD caregivers' mental health challenges, identifying eight key stressors: 1) disruptive behaviors by care-recipient~\cite{swearer1988troublesome,kunik2003disruptive,desai2012behavioral}, 2) lack of support~\cite{schulz2004family, jennings2015unmet,park2018roles}, 3) low self-efficacy~\cite{tang2016effects,gallagher2011self,crellin2014self}, 4) emotional distress~\cite{aminzadeh2007emotional,caputo2021emotional,gallego2022feel}, 5) relationship tensions~\cite{ray2022impact,vroman2019impact}, 6) compassion fatigue~\cite{day2011compassion,day2014compassion,perez2022mindfulness}, 7) lack of self-care~\cite{waligora2019self,oliveira2019improving,wang2019impact}, and 8) burnout~\cite{alves2019burnout,truzzi2012burnout,takai2009experience}. 
% These themes were selected based on prior research highlighting the psychological burden and coping mechanisms of caregivers. 

% From these themes, we developed eight detailed scenarios that caregivers frequently encounter. 
% Each scenario focused on a particular challenge, such as managing a care recipient's aggressive behavior, struggling with the lack of emotional support, or feeling overwhelmed by caregiving responsibilities. 
For each of the above eight scenarios, we crafted descriptions to help participants understand the context, along with the specific tasks that encouraged them to engage with our AI chatbot---as presented in~\autoref{tab:scenario_descriptions}. 
\edit{During the interview, participants were presented with the full list of eight caregiving scenarios and were first asked to select the one they found most relevant to their own caregiving experience. They then interacted with \carey{} by posing a query related to the selected scenario.}
% They would then choose a scenario and interact with \carey{} with a query.
% \edit{After completing the interaction and follow-up interview questions, participants were asked to select another scenario they could relate to and repeat the process.}
\edit{After completing the interaction and discussing their reactions, perceptions, and thought processes during the interaction, participants were invited to select another scenario they could relate to and repeat the process.}
\edit{This iterative procedure continued over the one-hour interview session, allowing participants to engage with multiple scenarios as time permitted, while ensuring that each interaction was grounded in situations they personally recognized rather than unfamiliar or hypothetical contexts.}
The number of scenarios each participant engaged with varied depending on their interaction pace, within the stipulated total interview time of one hour. 
% During the interviews, the eight scenarios were presented in random order, and participants were asked to choose one that they related with their own experience.
% whether they found each scenario relevant to their own experience. 
% If a participant related to a scenario or had insights to share, 
% They would then choose a scenario and interact with \carey{} with a query.
% to explore potential responses and coping strategies. 
% Otherwise, they skipped that specific scenario. 
% The number of scenarios each participant engaged with varied depending on their interaction pace, within the stipulated total interview time of one hour. 
% By structuring the scenarios around established caregiving challenges, we aimed to assess how AI-driven chatbots align with caregivers' needs, explore gaps in AI support, and identify future design improvements. 
%In the next section, we elaborate on the development of \carey{}.

\begin{table}[t!]
\centering
\sffamily
\footnotesize
\caption{Descriptions and tasks for each scenario in the study.}
\label{tab:scenario_descriptions}
\setlength{\tabcolsep}{3pt}
\begin{tabular}{p{0.11\columnwidth}p{0.43\columnwidth}p{0.41\columnwidth}}
% \toprule
\textbf{Scenario} & \textbf{Description} & \textbf{Task} \\
\toprule
\textbf{Disruptive behaviors by \hfill \break care-recipient} & Your care recipient has disruptive behaviors, such as \hfill \break aggressive behaviors (verbal outbursts, physical aggression), restlessness, and heightened irritability over minor triggers. You are seeking to better understand your care \hfill \break recipient’s behavior. & Start chatting with \carey{} about strategies for handling disruptive behaviors by your care-recipient. Ask for advice on managing verbal outbursts, physical aggression, and restlessness. Seek tips on identifying triggers and creating a calming environment. \\
\rowcollight \textbf{Lack of \hfill \break support} & You're facing challenges due to a lack of support systems to help manage your mental health. You have difficulties getting help from senior centers, professionals, and services like home health aides and respite care. & Chat with \carey{} about handling caregiving with limited support. Ask for tips on finding resources and \hfill \break support systems. \\
\textbf{Low \hfill \break self-efficacy} & You’re experiencing doubt in your ability to manage \hfill \break caregiving challenges. This lack of confidence affects your motivation and resilience. & Start a chat with \carey{} to discuss strategies for boosting self-efficacy. Seek advice on building confidence, staying resilient, and developing coping mechanisms. \\
\rowcollight \textbf{Emotional distress} & You’re struggling with emotional wellness, finding it difficult to stay positive and manage stress related to caregiving, including depression, guilt, and embarrassment. & Start a chat with \carey{} about strategies for maintaining emotional wellness. Ask for tips on managing guilt, \hfill \break depression, and other challenging emotions while \hfill \break staying resilient. \\
\textbf{Relationship tensions} & You’re struggling to manage family relationships due to the uneven distribution of caregiving duties, causing stress, \hfill \break tension, and conflict. & Start a chat with \carey{} about managing family \hfill \break relationships while caregiving. Ask for advice on handling conflicts, communicating with family, and sharing \hfill \break responsibilities more equally. \\
\rowcollight \textbf{Compassion fatigue} & You’re feeling emotionally drained and struggling to stay \hfill \break empathetic, especially as your care recipient's \hfill \break condition worsens. & Start a chat with \carey{} about handling compassion \hfill \break fatigue. Ask for tips on recharging emotionally, coping with burnout, and rekindling empathy. \\
\textbf{Lack of \hfill \break self-care} & You’re struggling to find time for self-care, feeling \hfill \break overwhelmed with caregiving responsibilities, and lacking personal space or privacy. & Start a chat with \carey{} about making time for self-care. Ask for tips on balancing responsibilities, setting \hfill \break boundaries, and incorporating self-care into daily life. \\
\rowcollight \textbf{Burnout} & You're feeling overwhelmed and exhausted, showing signs of burnout from caregiving demands. Persistent stress is \hfill \break impacting your emotional wellbeing and \hfill \break sense of accomplishment. & Start a chat with \carey{} about coping with caregiver burnout. Ask for advice on managing stress, improving emotional resilience, and regaining a sense of control. \\
% \bottomrule
\end{tabular}
\end{table}

\subsubsection{Development of \carey{}}
To develop \carey{}, we used the Flask web framework for its simplicity and flexibility in web applications, enabling seamless integration with the OpenAI API~\cite{openai2024models}.
% , which is the back-end of \carey{}.
% 's conversational capabilities. Carey 
\carey{} was implemented within a Python virtual environment, ensuring secure management of the OpenAI API key. 
For the back end, we selected GPT-4o---the state-of-the-art generative AI model at the time of this research---known for its advanced natural language capabilities and optimized response times.
% due to its advanced natural language processing capabilities and the optimal response time of this research. 
The API was incorporated into the Flask application with careful attention to constructing messages that defined the chatbot's role, the specific caregiving scenario, and the ongoing chat history. 

The chatbot function processed scenario descriptions, chat logs, and prompts
% (e.g., ``You are a peer-supporter chatbot designed to support the mental wellbeing of caregivers of Alzheimer's Disease and Related Dementias'') 
to generate relevant and contextually appropriate responses. 
We experimented with several prompts and caregiver queries (gathered from the body of work on identifying the scenarios as well as from online communities~\cite{saha2025ai,kaliappan2025online}) to arrive at a final prompt which led to generating most relevant responses---``You are a peer-supporter chatbot designed to support the mental wellbeing of caregivers of Alzheimer's Disease and Related Dementias. The caregiver wants to seek your advice and support related to the [S], which is described as [D]''---here [S] refers to the particular scenario and [D] consists of the scenario description (ref:~\autoref{tab:scenario_descriptions}).

\begin{figure}[t]
    \centering
    \begin{subfigure}[t]{0.44\columnwidth}
        \centering
        \includegraphics[width=\columnwidth]{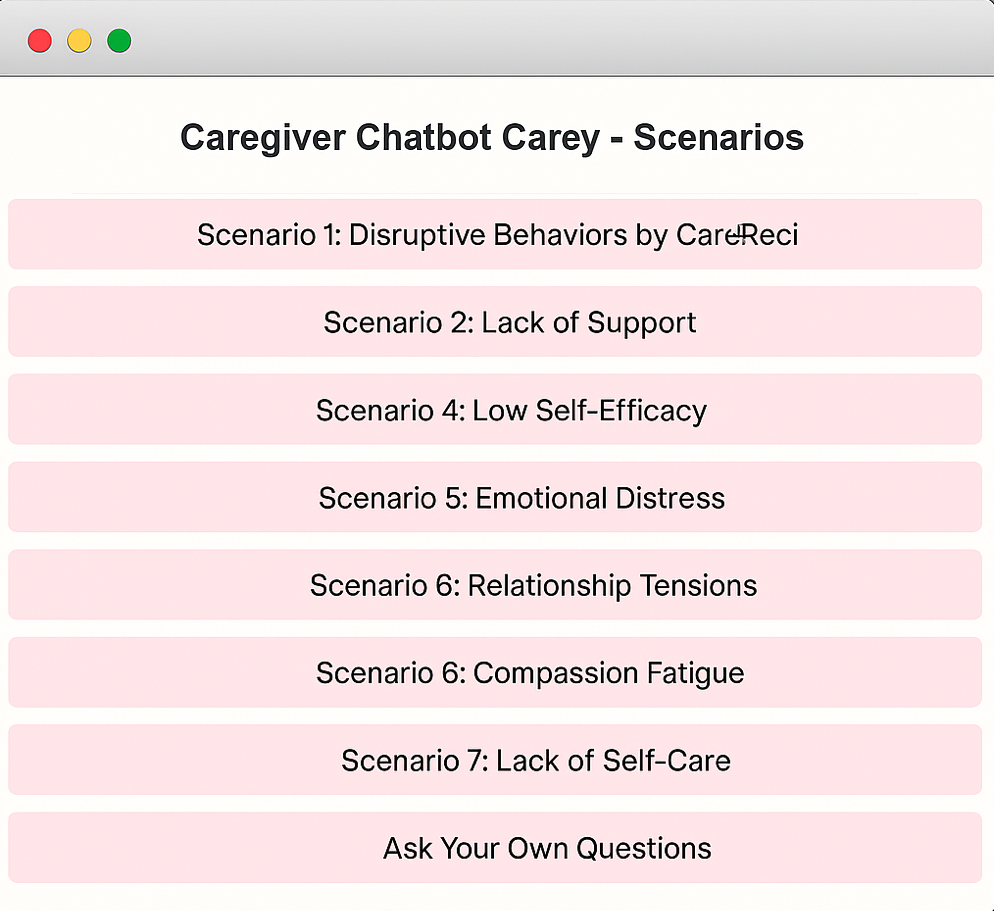}
        \caption{Scenario Page}
        \label{fig:login}
    \end{subfigure}\hfill
    % \vspace{1em} 
    \begin{subfigure}[t]{0.53\columnwidth}
        \centering
        \includegraphics[width=\columnwidth]{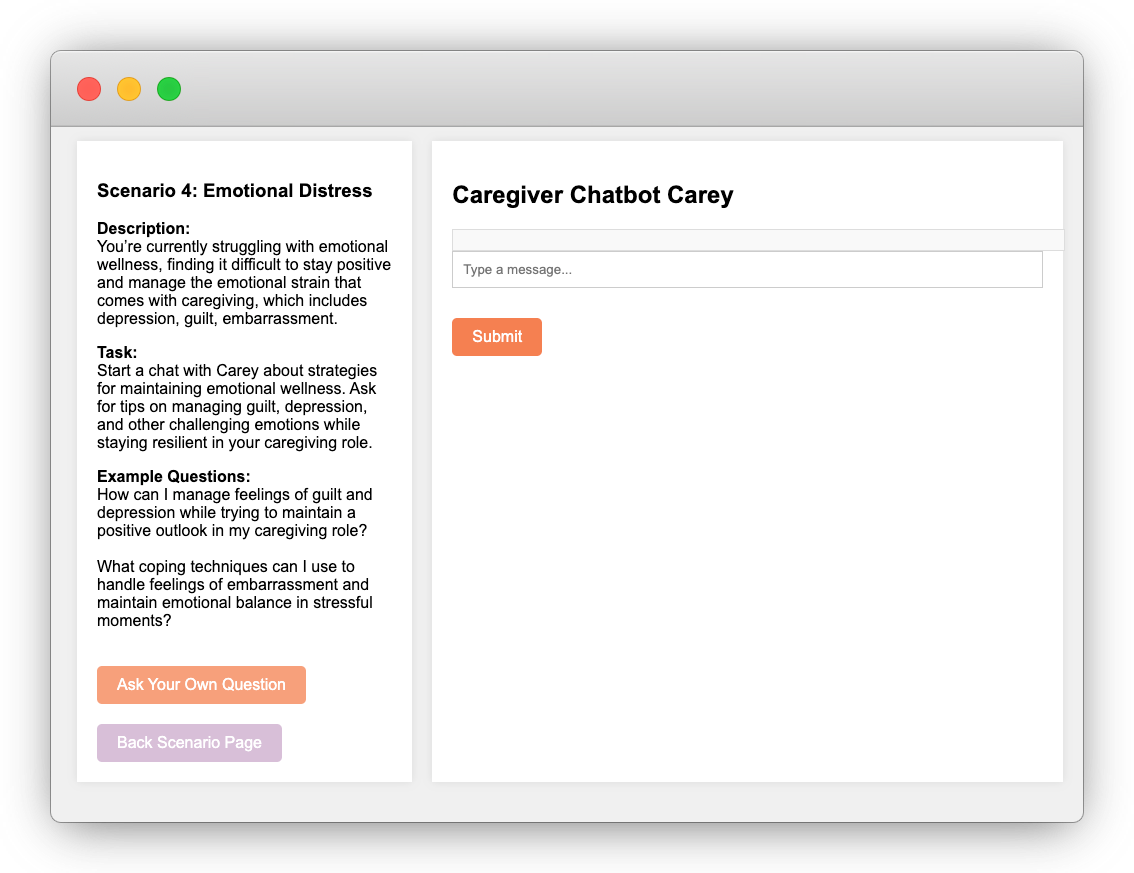}
        \caption{Chat Page}
        \label{fig:chat}
    \end{subfigure}
    \caption{Screenshots of the scenario and chat page of our prototype chatbot, \carey{}.}
    \Description[figure]{description of figure}
    \label{fig:Carey}
\end{figure}

\carey{}'s front-end interface followed a minimalist and simple design for clarity and ease of use, with the left part displaying the scenario number, title, description, task, and example questions, and the right part facilitated participant interactions (see~\autoref{fig:Carey}). 
The participant had options to go back to the scenario page and navigate between different scenarios. 
The participants' chatlogs and scenario selections were automatically stored on a secure server, which were later used for analysis alongside interview transcripts. 

\subsubsection{Safety Measures}
Given the emotional and psychological demands of caregiving for individuals with AD/ADRD, we implemented several safety measures to ensure participant wellbeing throughout the study. 
% Our research team included HCI researchers expertise in digital health, as well as advisors with experience in dementia caregiving and mental health. 
We carefully considered potential risks related to discussing caregiving challenges and emotional distress, taking proactive steps to minimize any negative impact on participants.

To ensure a safe interaction with AI chatbot \carey{}, we rigorously tested its responses across multiple caregiving scenarios before participant engagement. These scenarios were derived from existing literature on caregiver support needs and emotional wellbeing~\cite{randall2018engaging,lederman2019support, kim2024opportunities}. 
We rigorously tested several queries to ensure that the responses did not contain distressing or misinformation content. 
% Any responses that could lead to distress or misinformation were flagged and refined to align with ethical and supportive conversational standards. 
% \koustuv{@Melissa, could you add an example to elaborate on this point?} \melissa{Sorry, I don't have specific examples for this, I think we just test carey by reviewing that the responses were appropriate for AD/ADRD questions, maybe we can remove this sentence.}

During the study, one participant showed signs of emotional distress while discussing the caregiving experience. The interviewer provided space for them to process their emotions, offered a break, and reiterated the availability of support services. No other participants reported distress following the interviews.

\subsection{Interview Procedure}
We conducted semi-structured interviews with caregivers to explore their experiences and mental health throughout the caregiving journey. These interviews were conducted via video calls (Teams) and lasted approximately 60 minutes. The research team alternated between interviewing and note-taking during the sessions, which were recorded for research purposes. At the beginning of the interview, we introduced the study by explaining its purpose---exploring caregiving opinions on using an AI-based chatbot, \carey{}, for mental wellbeing support. \edit{We informed participants that all their interactions and communication with Carey during the interview would be kept confidential and used solely for research purposes.}
We aimed to investigate the needs, desires, and concerns of an AI chatbot catering to the mental health of AD/ADRD family caregivers. 
We informed participants that \carey{} was designed based on prior research that identified eight major causes of caregivers' mental health concerns. 
Before proceeding, we addressed any questions they had and sought permission to record the session. 
\edit{In our study, caregivers interacted with \carey{} during scheduled one-hour interview sessions, selecting scenarios that reflected their lived experiences. While these interactions occurred in a controlled setting, our interviews revealed that despite their heavy workloads and limited free time, many caregivers expressed a strong desire for mental health support like \carey{}. They emphasized that the flexibility and immediacy of an AI chatbot would allow them to seek emotional relief and practical advice during short breaks, late at night, or while multitasking. This suggests that caregivers value accessible, low-burden tools that can realistically fit into their demanding daily routines.}
The interview was structured into three main sections. First, we asked participants about their general caregiving and mental health experiences. We inquired about their care recipients, the duration of their caregiving experience, the impact on their mental health, specific mental health challenges they had encountered, and which aspects of caregiving they found particularly stressful or overwhelming~\cite{blum2010family}. In the second section, we focused on participants' experiences with AI-based chatbots. We asked whether they had previously used chatbots such as OpenAI's ChatGPT, Google's Gemini, or Microsoft's Bing, their motivations for doing so, their first impressions, and whether they had ever used chatbots for mental health support. If applicable, we asked them to describe their experiences using AI for mental health purposes, including the effectiveness and limitations of such tools. Finally, in the third section, participants engaged in direct interaction with \carey{} by selecting scenarios that resonated with their caregiving experiences. During these interactions, we encouraged them to think aloud—sharing their intentions behind each prompt and their reactions to \carey{}'s responses. Wherever applicable, we also prompted them to elaborate on their thoughts and personal experiences.

\edit{The eight scenarios were drawn from prior literature, and during interviews, participants selected the one most relevant to their lived experience, ensuring their reflections were grounded in real-life contexts rather than hypothetical situations}

% \subsection{Design of Scenarios}
% \koustuv{This subsection needs to be written describing how did we choose these scenarios and what each of the scenarios are about. You can get content from the CSCW paper. }
% \melissa{We have 3.1.1 "Scenario Design"}

\subsection{Data Analysis}
After the interviews, we used the default transcription feature on Teams for all interviews conducted. 
\edit{All interview recordings were carefully reviewed: the automatically generated transcripts from Teams were manually corrected by the research team through repeated listening to the recordings to ensure accuracy before analysis.} The recordings were anonymized by redacting any identifiable data such as personal names and locations. The data set was then treated as a corpus for comprehensive analysis. 
We analyzed our data, including transcriptions from the interview recordings, notes taken during the interviews, and chat history between the participants and \carey{}, using reflexive thematic analysis~\cite{braun2019reflecting}. 
Four co-authors participated in reviewing the transcripts and engaged in an interactive process of open coding, where codes were grouped, initial subthemes were identified, and subthemes were refined into higher-level themes. 

To further elaborate on our process, four co-authors were involved in the open coding process. 
The first author led the coding effort, completing the majority of coding on raw interview transcripts, while senior co-authors (who have extensive experience in qualitative research) provided feedback and suggestions during hybrid (in-person and screen-sharing-based) co-working sessions. 
To ensure coherence in the analysis, we carefully reviewed and refined the themes through an iterative process. This involved merging related themes into broader categories, separating overlapping themes into distinct categories, and discarding themes that were not directly relevant to our core research questions. 
While we allowed flexibility in our coding and theme development, we applied thematic analysis~\cite{braun2019reflecting} as our primary methodological approach, guided by prior research on social support in online dementia communities~\cite{johnson2022s}. 
We initially obtained 420 codes, which were grouped into 24 lower-\edit{level} subthemes and five higher-level themes that aligned with our research question.
% then through systemic grouping and refinement, these codes were consolidated into 24 lower-level subthemes and five higher-level themes that aligned with our research questions. 
% We also conducted thematic analysis~\cite{braun2019reflecting} on the chat history between \carey{} and participants, characterizing and understanding their conversational patterns and interactions.
% Ultimately, we determined the final thematic structure that corresponded to each research question through collaborative discussion and consensus among all co-authors.  

% \koustuv{@melissa, This needs to be elaborated further. Please take a look at inductive coding and thematic analysis as explained in the CSCW paper.} \melissa{done}

\subsection{Privacy, Ethics, and Reflexivity}
Our study was approved by the Institutional Review Boards (IRBs) at the researchers' institutions. 
Given the sensitive nature of caregiving experiences, we implemented several ethical and privacy considerations to protect participants' wellbeing and confidentiality. 
Each participant was assigned a unique participant ID to anonymize their responses, and all personally identifiable information was removed during transcripts and analysis. 
Throughout the interviews, we adopted a respectful and empathetic approach to discussing caregiving experiences. 
We used language that prioritized the caregiver’s perspective and emotional well-being, referring to their loved ones as ``care recipients'' rather than clinical terms such as ``AD/ADRD patients.'' 
During the interviews, we closely monitored participants' emotional states, paying attention to verbal and nonverbal cues. If a participant became emotional while sharing their experiences, we paused the session, offered them the option to take a break, and reminded them that they could discontinue at any time without consequences.

Our research team comprises researchers holding diverse gender, racial, and cultural backgrounds, including
people of color and immigrants with interdisciplinary expertise in HCI, digital health, AI ethics, and psychology as well as experience in studying caregiving technologies and social support. 
The team includes a clinical psychologist with experience in mental health and caregiver wellbeing, as well as researchers with personal caregiving experiences—though not specifically related to AD/ADRD. 
% Some researchers on the team have personal caregiving experiences, though not specifically related to AD/ADRD. 
Although we have taken great care to faithfully analyze and represent participants' perspectives, we acknowledge that our positionality as researchers and, in some cases, as caregivers may shape our interpretations. 
We remain committed to presenting participants’ experiences authentically and recognizing the diverse and complex nature of caregiving.

%% file: 4.1results_interactions.tex
\section{Results}
Overall, our analysis revealed predominantly favorable opinions of \carey{} among AD/ADRD caregivers ($N$=16). 
Nearly all participants (\edit{15 out of 16 participants}) described \carey{} as ``helpful'' or ``very helpful'' in addressing caregiver-related mental health challenges. \edit{The remaining participant expressed skepticism toward AI-based support and indicated a preference for speaking with a human instead.}
% Nearly all participants (\edit{15 out of 16 participants})
% (93.75\%) 
% described the prototype \carey{} as ``helpful'' or ``very helpful'' in addressing caregiver-related mental health challenges. 
First, our scenario-based analysis revealed that caregivers most frequently engaged with scenarios they found personally relevant and emotionally resonant. 
These usage patterns reflect caregivers' prioritization of mental health concerns that are often overlooked in general-purpose wellbeing tools. 
Second, our reflexive thematic analysis~\cite{braun2019reflecting} revealed that although caregivers appreciated \carey{}'s accessibility, emotional validation, and neutrality, they also identified key limitations in its current design. 
These insights underscore caregivers' desire for AI tools that are not only empathetic and nonjudgmental but also proactive, trustworthy, and attuned to the complex realities of caregiving.
In this section, we elaborate on our findings.

% \melissa{analysis about how these findings are unique for caregivers, vs, any general purpose mental health chatbot}

% \subsection{Qualitative Patterns in Caregiver--\carey{} Conversations}
\subsection{Characterizing and Understanding Caregiver--\carey{} Conversations}\label{sec:result_chat}
% \koustuv{can also include the quant/descriptive analysis. How many interactions per scenario. How long did each conversations last. } \melissa{duration data is hard to find, some are missing in the transcripts.}
% Our scenario-based analysis revealed distinct patterns in how caregivers engaged with \carey{} across different caregiving challenges. 
Participants in our study interacted with \carey{} through eight theory-driven scenarios known to be challenging for caregivers' mental health~\cite{van2015patients}.
% \koustuv{We need to correct these calculations based on how many participants chose each interactions}
% Our interviews with 16 participants resulted in a total of 93 scenario-based interactions, at an average of 5.85 interactions per participant. 
They prioritized interacting with scenarios that appealed the most to their lives---at an average of 3.25 scenarios per participant, and
% We conducted interviews with 16 caregivers, resulting in 93 total caregiver-Carey interactions across different scenarios.
% The 16 part
\autoref{table:mentalwellbeing} presents the frequency of occurrence of each scenario.
% with which each scenario was selected and the distribution of interactions across scenarios, highlighting those most closely aligned with participants' lived experiences. 
The scenario most chosen was emotional distress (by 12 participants), followed by disruptive behaviors by care-recipients (8 participants) and burnout (7 participants).
We describe the observations per scenario below:

\begin{table}[t!]
\centering
\sffamily
\footnotesize
% \caption{Mean Responses for Each Question (Numerical)}
\caption{Summary of participants' interactions with \carey{}.}
\label{table:mentalwellbeing}
\begin{tabular}{lll}
% \textbf{Question} & \textbf{Mean} & \textbf{Std. Dev.} & \textbf{Distribution}\melissa{not accurate}\\
\textbf{Scenario} & \textbf{Num. Participants} & \textbf{Num. Interactions}\\
\toprule
% \rowcollight \multicolumn{4}{c}{\textbf{Activities}}\\
Disruptive behaviors by care-recipient & \nuLRbar{.8}8 & \nuLRbar{1.5}15 \\
\rowcollight Lack of support & \nuLRbar{.2}2 & \nuLRbar{.4}4\\
Low self-efficacy & \nuLRbar{.3}3 & \nuLRbar{.6}6 \\
\rowcollight Emotional distress & \nuLRbar{1.2}12 & \nuLRbar{2.8}28 \\
Relationship tensions & \nuLRbar{.5}5 & \nuLRbar{1.1}11 \\
\rowcollight Compassion fatigue & \nuLRbar{.4}4 & \nuLRbar{.9}9 \\
Lack of self-care & \nuLRbar{.5}5 & \nuLRbar{.8}8 \\
\rowcollight Burnout & \nuLRbar{.7}7 & \nuLRbar{1.2}12 \\
% \bottomrule
\end{tabular}
\end{table}

% \begin{figure}[H]
%     \centering
%     \includegraphics[width=0.8\textwidth]{figures/horizontal_side_by_side_clean (3).pdf}
%     \caption{Need Captions}
%     \label{fig:left_plot}
% \end{figure}

\para{Scenario 1: Disruptive behaviors by care-recipients (\n{N}=8).} 
% generated 15 interactions (16.1\%) focused on managing challenging care recipient behaviors. 
Caregivers sought strategies for handling care refusal situations, improving care-recipients' receptiveness to suggestions, managing verbal aggression, navigating public perception of inappropriate comments, and calming agitated care-recipients during outbursts. 
Some common questions included---``How do I calm my recipient when they are aggressive?'' (P12) and ``How can we manage inappropriate comments by patients?'' (P03). 
\carey{} responded with de-escalation techniques, person-centered validation approaches, and practical recommendations for explaining AD/ADRD-related behaviors to others.
% \koustuv{How did the participants respond to these conversations?}
% \melissa{
P3 noted that \carey{}'s advice aligned closely with the guidance they had previously received from an AD/ADRD caregiving support group they participated in. 
They found the responses helpful and expressed interest in continuing to engage with \carey{} for future support.

\para{Scenario 2: Lack of Support (\n{N}=2).} 
% produced 4 interactions (4.3\%) centered on systemic barriers to formal assistance. 
Although this scenario was only chosen by two participants, it reflected interactions on frustration with institutional and social support. 
% This low-frequency but critical scenario reflected frustration with institutional aid. 
For instance, P1 expressed, ``I am not getting enough help from the Veterans organization. How do I navigate through this situation?'' 
\carey{} validated P1's struggle, suggested persistence, and offered guidance on alternative advocacy options and caregiver networks, highlighting the limitations of AI when institutional change is needed.

\para{Scenario 3: Low Self-Efficacy (\n{N}=3).} 
% yielded 6 interactions (6.5\%) reflecting caregivers' uncertainty in decision-making processes. 
In this scenario, caregivers asked questions like, ``What tips are available to help me decide if treatments are the correct thing to do?'' (P13) and ``What practical tips can help me feel more capable and confident when facing new caregiving challenges?'' (P1). 
\carey{} responded with frameworks for shared decision-making, conflict resolution strategies for family discussions, and permission-giving language that emphasized the connection between caregiver wellbeing and care quality.

\para{Scenario 4: Emotional Distress (\n{N}=12).} 
% emerged as the most prevalent concern, generating 29 interactions (30.1\%) across diverse emotional challenges. 
In this scenario, example questions consisted of ``How do I manage my depression while taking care of my dad?'' (P3) and ``How do I deal with feeling low emotionally when caregiving seems endless?'' (P7). 
\carey{} provided validation-first responses, followed by reflective prompts and coping suggestions such as journaling, breathwork, and emotional boundary-setting.
% ---positioning this scenario as a core emotional inflection point.

\para{Scenario 5: Relationship Tensions (\n{N}=5).} 
% produced 11 interactions (11.8\%) focused on strained interpersonal dynamics. 
In this scenario, participants voiced challenges like, ``How do I get closer to my family when I care for my dad?'' (P4) and ``How can I talk to my siblings about their lack of involvement?'' (P1). 
\carey{} emphasized empathetic communication, boundary-setting, and collaborative planning, often providing sample scripts to reduce confrontation while advocating for shared responsibility.

\para{Scenario 6: Compassion Fatigue (\n{N}=4).} 
% generated 9 interactions (9.7\%) focused on emotional depletion from prolonged caregiving. 
Participants expressed questions regarding compassion fatigue and emotional withdrawal, ``I feel blank when I think about my life---how do I fix that?'' (P8) and ``How can I take care of my emotional needs while caring for someone who won't address theirs?'' (P14). 
% revealed deeper emotional withdrawal. 
\carey{} acknowledged the cumulative toll of caregiving and recommended micro-restoration practices, re-framing caregiver fatigue as a valid and treatable experience tied to prolonged emotional labor.

\para{Scenario 7: Lack of Self-Care (\n{N}=5).} 
% yielded 8 interactions (8.6\%) addressing caregivers' struggles to prioritize their own wellbeing. 
Questions in this scenario included, ``How do I balance caregiving with personal time?'' (P16) and ``What small self-care activities can I actually do between all my responsibilities?'' (P10). 
\carey{} offered actionable self-care strategies grounded in feasibility, such as incorporating mindful breathing into daily routines. 
These recommendations align with micropractices that accommodate time constraints while supporting gradual restoration.
% These micropractices respected caregivers' time constraints while promoting incremental restoration.

\para{Scenario 8: Burnout (\n{N}=7).}
% produced 12 interactions (12.9\%) reflecting caregivers in crisis states. 
This scenario captured caregivers expressing ambivalence or emotional collapse, ``I love her, but I'm exhausted. I don't want her to die, but I'm relieved of the thought'' (P15). 
Another participant asked, ``How do I manage work and caregiving when both demand everything from me?'' (P11). 
\carey{} responded with layered support---validation, time-sensitive coping plans, and referrals to external resources---recognizing burnout as both a psychological and practical challenge of caregiving.

\para{Own Questions (\n{N}=6).}
Six participants chose to engage with \carey{} by asking their own questions. These interactions reflected caregivers' personal interests, concerns, and curiosity, \edit{often extending into medically or genetically related topics.}
% inquiries that were medically or genetically oriented.}
%often extending into more medically or genetically specifically inquiries. 
For example, P3 asked a detailed question about hormone replacement therapy (HRT) in the context of carrying the Apolipoprotein E epsilon 4 (APOE4) gene variant: ``I have the APOE4 gene. Are there specific things people with this allele can do to reduce their risk of Alzheimer’s? What’s the latest research on hormone replacement therapy for women with APOE4?''. P3 noted that \carey{}'s responses surfaced new lines of inquiry they had not initially planned to explore: ``I didn't come in thinking I wanted to ask more about HRT and APOE4, but seeing what \carey{} provided prompted more questions, which I really liked.''. 
This shows how \carey{} can enable discovery-oriented dialogue by supporting relevant knowledge, especially in areas they might not have otherwise explored. 

% \koustuv{This needs to be elaborated.}

% \para{Cross-Scenario Insights} 
\para{Overall Insights.} Across all scenarios, participants used \carey{} to navigate both emotional validation and practical scaffolding. 
Emotional needs were particularly prominent, with participants often initiating conversations from moments of vulnerability, seeking reassurance, and meaning-making.
% Meanwhile, practical questions---
% \koustuv{Elaborate what do we mean by practical questions here.} though fewer in number---revealed a 
Participants also desired for trustworthy, situation-specific guidance that felt personalized and realistic within the constraints of caregiving.
%We observed that participants frequently commented on \carey{}'s tone, appreciating its non-judgmental, calm, and empathetic voice. 
\edit{We observed that participants frequently commented on \carey{}’s tone in written exchanges, noting its non-judgmental, calm, and empathetic communication style.}
However, the expectations varied. Some participants sought proactive guidance or more emotionally rich engagement, while others expressed skepticism about AI's ability to truly ``understand'' caregiving.

%% file: 4.3results_alignment.tex
% \subsection{Alignment and Misalignment of \carey{} with Caregivers' Needs}\label{sec:result_transcript}
\subsection{Needs and Expectations of Caregivers from \carey{}}\label{sec:result_align}
% Participants recognized both strengths and gaps in how \carey{} addressed their specific needs. 
Our analysis revealed that while participants generally found value in their interactions with \carey{}, they also recognized areas where the tool could more effectively address their caregiving challenges. 
In this section, we present key themes related to participants' needs and expectations, highlighting the tensions between the perceived benefits and concerns of interacting with \carey{}. \edit{In the findings below, we highlight participants' recognition of \carey{}'s strengths and gaps using \textbf{boldface}.}
% , which presents illustrative participant--\carey{} exchanges to ground the reported themes in concrete interactional examples.}
% but also recognized areas where it could better align with their caregiving challenges. 
% In this section, we describe the themes of participants' needs and expectations, and the tensions in their perceived benefits and concerns about \carey{}.
% map the alignment and misalignment between participants' expectations and \carey{}'s capabilities, highlighting key tensions between appreciation for its functionality and concerns about areas for improvement.

% In this section, we map the alignment and misalignment between participants' expectations and \carey{}'s capabilities, highlighting key tensions between appreciation for its functionality and concerns about areas for improvement.

\begin{table}[t]
\footnotesize
\sffamily
\centering
\caption{\edit{Example excerpts of caregiver--\carey{} interactions across scenarios.}}
\label{tab:carey_example_interactions}

\edit{%
\begin{tabular}{p{0.22\linewidth} p{0.73\linewidth}}
\textbf{Context} & \textbf{Excerpt} \\
\toprule
\textbf{P1, Scenario~2} \newline
(AI chatbots as a source of on-demand information access.) &
\textbf{Caregiver.} I am not getting enough help from the Veterans organization. How do I navigate through this situation?

\vspace{0.4em}

\textbf{\carey{}.} I’m sorry you’re having difficulty getting the support you need. One option is to contact a Veterans Service Officer (VSO)—they’re trained to advocate on your behalf and help you navigate available services. You may also find it helpful to clarify the specific support you’re requesting and follow up regularly, since delays and lost paperwork are common. I know this process can be frustrating, and you deserve reliable support as a caregiver. \\
\hdashline
\rowcollight \textbf{P10, Scenario~4} \newline
(AI chatbots for emotional support) &
\textbf{Caregiver.} I'm afraid that my wife is going to die. Alzheimer's is terminal. I'm too young to be a widower.

\vspace{0.4em}

\textbf{\carey{}.} I'm really sorry to hear that you're feeling this way. It's completely understandable to be afraid of losing your wife in the context of Alzheimer's. Your feelings of fear, sadness, and anxiety are valid, and you don't have to face them alone. Talking with trusted people or a therapist can help lighten the emotional load. Remember, you're doing an incredibly challenging job, and seeking support for yourself is an important and courageous step. \\
\hdashline
\textbf{P9, Scenario~1} \newline
(AI chatbots as a ``safe space'' for disclosure) &
\textbf{Caregiver.} How do I stop reacting to repetitive behavior?

\vspace{0.4em}

\textbf{\carey{}.} It's completely understandable to feel overwhelmed by repetitive behaviors, especially in caregiving situations. One thing that can help is taking a moment to pause and breathe before responding, so you have space to ground yourself. Gently redirecting your loved one’s attention, establishing a predictable routine, and minimizing possible stressors in the environment can also make these moments easier to navigate. Remember, it's natural to feel a range of emotions while caregiving, and being kind to yourself is just as important as the care you provide. \\
\hdashline
\rowcollight \textbf{P16, Scenario~4} \newline
(AI chatbots' personalization capabilities) &
\textbf{Caregiver.} What steps can I take when I'm overwhelmed with caregiving and need support?

\vspace{0.4em}

\textbf{\carey{}.} It sounds like caring for your mom has been overwhelming for you, and that feeling is completely valid. You’re carrying a lot right now, and you don’t have to manage it alone—reaching out to family, other caregivers, or short-term respite care could help. Even small breaks for yourself can make a difference. \\

\end{tabular}
} % ← end \edit
\end{table}

\subsubsection{AI chatbots as a source of on-demand information access}
\label{sec:result_align_1}

\textbf{Participants appreciated \carey{}'s potential as a tool for on-demand practical guidance, emphasizing the need for credible and contextual information.}
\edit{For example, P1's interaction in Scenario 2 (\autoref{tab:carey_example_interactions}) shows how caregivers may seek immediate and actionable guidance when navigating fragmented institutional support systems, such as seeking assistance from the veterans' organization.}, 
Participants responded positively to \carey{}'s ability as a source of immediate information access for diverse caregiving scenarios, e.g.,: P16 commented, ``[The information provided by \carey{}] is practical and useful.''
% caregivers valued \carey{}'s ability to provide immediate, actionable guidance when navigating fragmented institutional support systems, such as seeking assistance from veterans' organizations (\autoref {tab:carey_example_interactions}).}

% In particular, participants appreciated \carey{}'s ability to provide actionable guidance tailored to their caregiving scenario, e.g.,:

% \begin{quote}
% \small
% ``[The information provided by \carey{}] is practical and useful.''---P16
% \end{quote}

At the same time, participants also expressed concerns about the verifiability and trustworthiness of the information provided. 
% Unlike general mental health chatbots, where users may seek emotional validation, 
In fact, the AD/ADRD caregivers may need evidence-based guidance for complex care decisions. For example, P3 expressed a preference for information certified directly from trusted sources:

\begin{quote}
\small
``Personally, I just know the concept is that it pulls data from a variety of sources. But I would always want to go to an original source or a doctor, or the Alzheimer's website. I'd be worried about it pulling dirty data, like I'd be worried about it pulling information that's not accurate.''---P3
\end{quote}

This highlights a fundamental tension---although \carey{} can aggregate information from multiple sources, caregivers fear potential misinformation that could impact critical care decisions. Beyond trustworthiness concerns, participants also identified issues with information quality and relevance. For example:

\begin{quote}
\small
``It seems like recycled content from the previous questions that is only marginally relevant and useful''---P10
\end{quote}

It indicates frustration with repetitive responses lacking novel insights or contextual adaptation---this suggests that caregivers also expect content that demonstrates an understanding of specific caregiving context and avoid generic, templated responses that fail to address their evolving needs. 
\edit{These concerns reflect the uniquely high-stakes nature of caregiving, where informational inaccuracies carry greater consequences than in general wellbeing chatbots. Caregivers were not simply seeking quick answers---they expected \carey{} to demonstrate contextual intelligence and incorporate evidence-based sources. This underscores the need for tools that provide not only fluent responses but also verifiable, personalized, and dynamically updated information grounded in caregivers’ real-world situations.}

\subsubsection{AI chatbots as a ``safe space'' for disclosure}
% \melissa{example response from chat history}
% \begin{quote}

% \edit{\textbf{Question (P9, Scenario 1).} How do I stop reacting to repetitive behavior?}

% \edit{\textbf{\carey{}.} ``It's completely understandable to feel overwhelmed by repetitive 
% behaviors, especially in caregiving situations. [...] One thing that can help is taking a moment to pause and breathe before responding, so you have space to ground yourself. Gently redirecting your loved one’s attention, establishing a predictable routine, and minimizing possible stressors in the environment can also make these moments easier to navigate. [...] Remember, it's natural to feel a range of emotions while caregiving, and being 
% kind to yourself is just as important as the care you provide.'' }
% \end{quote}

\label{sec:result_align_2}
\textbf{Participants valued \carey{} as a judgment-free, always-available outlet for emotional venting, but desired more natural conversational reciprocity.}
They described \carey{} as a space where they could openly process emotions---such as guilt, frustration, or exhaustion---without the fear of judgment. 
This sense of psychological safety was essential, given that societal norms often discourage caregivers from voicing emotional strain.
\edit{For example, in P9's interaction in Scenario 1 (\autoref{tab:carey_example_interactions}), P9 openly reflected on frustration with repetitive behaviors without fear of social judgment.}
Much like journaling or expressive writing, \carey{} offered participants a private, stigma-free outlet for candid self-reflection and disclosure---particularly valuable when caregivers face complicated decisions regarding behavioral management or long-term planning, as P4 expressed:

\begin{quote}
\small
``It's not going to judge you, with humans, sometimes you feel like people see everything you do, and at some points, it might feel like, ``\textit{OK, you're way too emotional.}'' But with AI, it doesn't judge you like that.''---P4
\end{quote}

Therefore, the AI's neutral stance allowed participants to articulate their thoughts that they might otherwise suppress for fear of disappointing others or being perceived negatively. Further, P1 emphasized that AI provided the freedom to express without social consequences:

\begin{quote}
\small
``Sometimes it's hard to share emotions with another human being. But with an AI chatbot, you don't have to worry about being judged because it's just an AI---it doesn't have biases or personal opinions.''---P1
\end{quote}

Likewise, P9 noted that \carey{} enabled easier sharing of sensitive thoughts than real-world relationships shaped by power dynamics, 
% described how \carey{} made it easier to express sensitive thoughts than in real-world relationships shaped by power dynamics, 
\edit{``It actually is easier [to share] to \carey{} than to someone who may be my boss.''}

% \begin{quote}
% \small
% ``It actually is easier [to share] to \carey{} than to someone who may be my boss.'' ---P9
% \end{quote}

That said, participants expressed a desire for more natural conversational turn-taking---which was lacking in \carey{} in its current form. 
They found the interactions to be one-sided, placing the burden on caregivers to sustain engagement. P14 expressed frustration at needing explicit prompting: ``I told it to talk like a human, and it did. But I wish it would do that without me having to ask.''

% \begin{quote}
% \small
% ``I told it to talk like a human, and it did. But I wish it would do that without me having to ask.''---P14
% \end{quote}
Therefore, \carey{}'s reactive conversational approach fell short of creating the dynamic and back-and-forth rhythm that typically exists in supportive human interactions, as P3 noted:

\begin{quote}
\small
``I wish \carey{} would ask me questions too. That's what I like in a real conversation---it keeps me engaged.''---P3. 
\end{quote}

\subsubsection{AI chatbots for emotional support}
% \begin{quote}
% \edit{\textbf{Question (P10, Secnario 4).} I'm afraid that my wife is going to die. Alz is terminal. I'm too young to be a widower.}

% \edit{\textbf{\carey{}.} ``I'm really sorry to hear that you're feeling this way. It's completely understandable to be afraid of losing your wife, especially in the context of Alzheimer's. [...] Your feelings of fear, sadness, and anxiety are valid, and you don't have to face them alone. Talking with trusted people or a therapist can help lighten the emotional load. [...] Remember, you're doing an incredibly challenging job, and seeking support for yourself is an important and courageous step.''}
% \end{quote}
\label{sec:result_align_3}
Related to the above, \textbf{participants recognized and acknowledged \carey{}'s ability to offer emotional support, but also desired a depth of human connections.}
Caregiving often involves deep emotional labor, including compassion fatigue, anticipatory grief, emotional exhaustion, and role-related guilt, as noted in prior work~\cite{werner2010subjective, day2011compassion,day2014compassion,perez2022mindfulness}. 
Our interviews revealed that \carey{} showed strengths in providing emotional support that resonated deeply with caregivers---supporting prior work on AI's capabilities in emotional support and empathy~\cite{cai2019hello, berretta2023defining}. 
In particular, participants valued \carey{}'s ability to validate their complex feelings, and offer empathetic, emotionally attuned responses---something participants identified as a core unmet need in their caregiving experience. 
For example, P8 expressed:

% \begin{quote}
% \small
% ``Being a caregiver, at times, you need support. You need someone to hear you, someone actually to understand how you feel.''---P8
% \end{quote}

% \koustuv{this should go into the next theme.}
Interestingly, participants also interpreted certain system behaviors---such as a pause between input and response---as signs of thoughtful engagement rather than technical delay. For example, P13 said: ````It felt like it was thinking. I asked a hard question, and it took a moment. That made it feel more intelligent.''
% \begin{quote}
% \small
% ``It felt like it was thinking. I asked a hard question, and it took a moment. That made it feel more intelligent.''---P13
% \end{quote}
% \edit{This suggests that caregivers attributed intentionality and emotional presence to interaction cues, even when those cues stemmed from technical latency.}
% Such attribution aligns with prior findings on relational agents and highlights how minimal design elements can be perceived as empathy or attentiveness.}
% However, \carey{}'s support often felt limited to isolated exchanges and lacked the dynamic, emotionally rich engagement that human-human supportive interactions provide. P3 remarked:

% \begin{quote}
% \small
% ``I bet it can't do this with following up like ``\textit{which of these are you more interested in?}'' Like something you know when you're in human conversation.''---P3
% \end{quote}

While some participants recognized the benefits of AI in offering low-risk emotional expression, they also voiced a clear preference for authentic, human interactions. P3 emphasized:

\begin{quote}
\small
``If I am going to interact with anything, I want it to be human. So for example, I'm very involved in my Alzheimer's support group. I find the human interaction very valuable.''---P3
\end{quote}

Therefore, we found a nuanced tension in participants appreciating AI's supportive qualities and maintaining a strong preference for human interactions. Beyond preference, some participants voiced ethical and emotional concerns about growing too attached to AI-based support. P15 reflected:

\begin{quote}
\small
``In general, I feel a bit suspicious of AI. I don't want it to replace human interaction, and honestly, I don't want to grow too fond of it.''---P15
\end{quote}
\edit{These concerns reflect deeper anxieties about dependency, authenticity, and the possibility of AI inadvertently reshaping caregivers’ social worlds. While caregivers welcomed immediate emotional relief from \carey{}, they were wary of emotional overreliance or the erosion of valued human support systems.}

\subsubsection{AI chatbots for crisis management}
\label{sec:result_align_4}
% \begin{quote}
% \edit{\textbf{Question (P3, Scenario 1).} How can we manage inappropriate or aggressive comments from my father?}

% \edit{\textbf{\carey{}.} ``When he makes an inappropriate comment, try acknowledging the feeling behind it rather than the words themselves—for example, ‘I can see you’re upset.’ Gently redirecting him to a comforting activity like music or a photo album can help de-escalate the moment. Identifying possible triggers and keeping the environment calm can also reduce the frequency of these episodes.’’}
% \end{quote}
Our interviews revealed that \textbf{although participants appreciated \carey{}'s ability to provide timely and round-the-clock interactions and support, it needs better crisis response.}
To begin with, \carey{}'s potential 24/7 availability strongly aligned with caregivers' need for on-demand assistance during unpredictable caregiving situations. 
\edit{In moments of anticipatory grief, such as P10's interaction in Scenario 4 (\autoref{tab:carey_example_interactions}), P10 appreciated Carey’s empathetic tone and emotional validation when confronting fears about disease progression and loss.}
Participants consistently valued the immediate accessibility of guidance that mitigated waiting times experienced with human support systems. 
This was particularly valuable given the irregular and often crisis-driven nature of AD/ADRD caregiving, where challenges frequently emerge during off-hours when professional support may be unavailable. 
This shows an AI's potential to address a critical vulnerability in the caregiving journey---isolation during moments of urgent need.
For example:

\begin{quote}
\small
``Using an AI chatbot gives you that 24/7 availability. They're always accessible and sometimes provide resources you didn't even know existed.''---P8
\end{quote}

This ``always available'' nature of \carey{} was considered to be ``emotionally stabilizing'' for multiple participants.
This contrasted with participants' experiences trying to contact healthcare professionals, family members, or support groups during moments of need, as P9 expressed:

\begin{quote}
\small
``Instead of waiting for responses from professionals who may be available, I get an immediate answer. It's a real-time interaction.''---P9
\end{quote}

\edit{Despite these strengths, caregivers expressed concerns about \carey{}'s ability to support time-sensitive or high-stakes situations. In particular, moments of delay raised uncertainty, 
% about how it would perform under more urgent conditions. 
such as P3 noted that although \carey{} eventually ``gave me the information I wanted, [the delay] made me wonder how it would handle something more serious or urgent.''
Such reflections highlight the importance of context-aware escalation mechanisms when caregivers face situations that may require immediate professional support.}

\subsubsection{AI chatbots' personalization capabilities}\label{sec:result_align_5}
% \begin{quote}
% \edit{\textbf{Question (P2, Scenario 4).} How can I stay afloat and manage my stress and guilt?}

% \edit{\textbf{\carey{}.} ``It’s understandable to feel emotionally drained while caregiving. Practicing self-care, mindfulness, journaling, or taking short breaks can help you manage stress. Connecting with support groups may also help you feel less alone.’’}
% \smallskip
% \textit{(Participant noted this felt generic and repetitive rather than tailored to their situation.)}
\textbf{A cross-cutting theme was that participants appreciated \carey{}'s personalization capabilities; however, they also sought a deeper depth of engagement for complex caregiving situations.}
In particular, they frequently described \carey{} as helpful in offering seemingly tailored responses that resonated with their situations. Many were surprised by how personally relevant the chatbot's suggestions felt. As P9 described:

\begin{quote}
\small
``I was surprised by [\carey{}'s responses]. It felt like it was a tailored, real-time self-help book or support group, which is especially [helpful as] caregiving needs [to increase].''---P9
\end{quote}

Participants pointed to small psycholinguistic cues---such as the use of personal pronouns and conversational tone---as contributing to a sense of individualized interaction. P2 noted, ``I just like [\carey{}] using ``I'' and being more personalized. It gradually resonates with me.''

% \begin{quote}
% \small
% ``I just like [\carey{}] using ``I'' and being more personalized. It gradually resonates with me.''---P2
% \end{quote}

% P13 further noted that \carey{} could adapt its general knowledge to address specific queries.

% \begin{quote}
% \small
% ``General responses, but it's specific to what I type.''---P13
% \end{quote}

Despite these strengths, caregivers consistently pointed to limitations in depth and contextual understanding. While initial interactions felt responsive, many participants found the personalization to be surface-level and not well attuned to the complexity of their caregiving situations. For instance, P8 remarked:

% \begin{quote}
% \small
% ``So I think this is not as personalized as it could actually be, probably because I just started chatting with it.''---P8
% \end{quote}

% Others expressed that responses often felt mismatched in tone, length, or emotional nuance. P10 explained:

% % \begin{quote}
% % \small
% ``I'm asking sophisticated questions--personal questions that I believe are complicated--but [\carey{}'s] answers are too long for what I need.''---P10
% % \end{quote}

Beyond functional assistance, participants emphasized the need for emotional support that felt attuned to the complexity of their caregiving roles. In high-stress moments, generic reassurance often fell short, underscoring the importance of deeper personalization. 
% This highlights an opportunity for future systems to move beyond surface-level responses and develop adaptive user models that reflect caregivers' evolving emotional and informational support needs. 
Participants also expressed a desire for AI tools to facilitate meaningful real-world connections---such as recommending peer support groups or strategies for carving out time for self-care. They valued support that acknowledged both their emotional challenges and their social isolation, offering not just empathy but concrete avenues for connection.
For example, P8 appreciated that the system did not just validate their loneliness but offered actionable strategies across multiple scenarios, such as taking regular breaks and engaging with caregiving communities:

\begin{quote}
\small
``[\carey{}] literally gave me ways what to do to: manage my feeling of loneliness and it actually gave a whole lot of scenarios I can connect with others. I can schedule regular breaks and join a caregiving community.''---P8
\end{quote}
% \edit{Caregivers further highlighted the importance of perceived personalization, with P16's interaction in Scenario 4 demonstrating how Carey’s acknowledgment of personal caregiving context contributed to feelings of being understood and supported (Table)~\ref{tab:carey_example_interactions}.}
This illustrates how effective AI support bridges emotional reassurance with practical guidance, helping caregivers feel both understood and \edit{empowered} through connections to real-world support networks.

% \subsubsection{Tensions around privacy}
\subsubsection{AI chatbots and perceptions of privacy}
\label{sec:result_align_6}
% \begin{quote}
% \edit{\textbf{Question (P10, Scenario 3).} I feel horrible. I'm so tired of looking after my wife. 
% I'm conflicted—I'm kind of looking forward to her death, but I don’t want her to die.}

% \edit{\textbf{\carey{}.} ``It’s completely understandable to feel overwhelmed and conflicted. Caregiving can take an enormous emotional toll. These thoughts don’t make you a bad person—they reflect the strain you’re under, not your character. 
% You deserve a space to express these feelings safely. Reaching out to a support group or a therapist can help you process them without judgment. You don’t have to carry this alone.''}
% \end{quote}

\textbf{Participants expressed complex and contrasting perspectives on privacy in their interactions with \carey{}, revealing a nuanced tension between perceived comfort and hesitation around disclosing to an AI.}
% in anonymity, and a hesitation around deeper disclosure.}
For many participants, \carey{}, being an AI agent created a reassuring sense of protection. 
Multiple participants described feeling less concerned about privacy because \carey{} lacked personal knowledge of them or their care recipients. 
For example, \edit{P15 appreciated a degree of privacy ``because \carey{} doesn't know me and my [care-recipient].''}
% \carey{} doesn't know my [care-recipient].''

% described that : ``My perception is that there is a degree of privacy here because \carey{} doesn't know me. \carey{} doesn't know my [care-recipient].''

% \begin{quote}
% \small
% ``My perception is that there is a degree of privacy here because \carey{} doesn't know me. \carey{} doesn't know my [care-recipient].''---P15
% \end{quote}

Likewise, several participants expressed preference to disclose sensitive information to an AI versus humans. P7 optimistically commented, ``AI, unlike humans cannot steal information,'' and P4 noted about being more open to an AI rather than a human:

% \begin{quote}
% \small
% ``I prefer going in depth with AI chatbots because I believe there is privacy.''---P4
% \end{quote}

% Similarly, P2 appreciated that \carey{} \textit{respected} personal boundary by not asking for unsolicited personal details: 

% \begin{quote}
% \small
% ``It doesn't ask anything about my personal life or anything to do with me personally [..] It doesn't have an effect on my personal private life. I am more concerned about people knowing me, my identity, and my questions.''---P2
% \end{quote}

% Similar preference in disclosing sensitive information to an AI versus humans was echoed by other participants---P7 optimistically commented that ``AI, unlike humans cannot steal information,'' and P4 commented:

% \begin{quote}
% \small
% ``I prefer going in depth with AI chatbots because I believe there is privacy.''---P4
% \end{quote}

% Further, the perceived anonymity enabled comfort in disclosing mental health struggles or seeking advice, e.g.,:

% \begin{quote}
% \small
% ``I feel comfortable enough to sufficiently describe my situation to Carey without identifying myself directly.''---P10
% \end{quote}

However, the above comfort was not universal. For instance, P15 was comfortable in sharing their own mental health struggles, but expressed caution in discussing about others (e.g., their care-recipient):
% revealing only limited information when it concerned others (e.g., their care-recipient):

% how much to reveal, particularly when discussing others (e.g., their care-recipient):

\begin{quote}
\small
``I don't mind talking about myself and my own mental health, but when it comes to [my care-recipient], I will protect their privacy 100\%.''---P15
\end{quote}

% Even those who felt safe acknowledged a threshold---often shaped by not knowing how data is stored, retrieved, or used.

% \begin{quote}
% \small
% ``At this point, I feel like the cat’s out of the bag." — P9
% \end{quote}

% Further, participants raised privacy considerations. 
% While they valued real-time assistance during critical situations, they expressed hesitation about sharing detailed personal information without a clear understanding of data handling practices. 

Similarly, P10 noted that they would share sensitive information under a veil of anonymity: ``I feel comfortable enough to sufficiently describe my situation to \carey{} without identifying myself directly.''
% As P10 noted, 
% \begin{quote}
% \small
% ``I feel comfortable enough to sufficiently describe my situation to \carey{} without identifying myself directly''---P10
% \end{quote}

% This suggests that anonymity was an important factor in conversations with the AI chatbot. 
Some participants expressed a sense of resignation regarding data security in the digital age, with P9 stating, ``At this point, I feel like the cat's out of the bag.''
% indicating broader concerns about information privacy in AI interactions. 
However, the desire for immediate support sometimes conflicted with privacy concerns, as more personalized crisis assistance would require sharing more detailed personal information. 
% This was particularly evident during discussions of behavioral management strategies requiring specific details about care recipients' conditions or when caregivers needed to describe emergency situations in detail to receive appropriate guidance.
% Notably, participants also struggled with the trade-off between privacy and personalized support. 
Some participants recognized the need to share more detailed and sensitive information during crises or when seeking behavior-specific guidance. 
This introduced a dilemma---while privacy was valued, there was also an urgent desire for actionable, tailored help, such as P1 shared, ``I'm comfortable sharing sensitive info with \carey{} to get a more specific answer.''

% \begin{quote}
% \small
% ``I'm comfortable sharing sensitive info with \carey{} to get a more specific answer.''---P1
% \end{quote}

Relatedly, P4 shared that they were comfortable sharing more personal information, but explicitly emphasized the expectation of protecting the privacy of the chat history---drawing comparison with therapist sessions:

\begin{quote}
\small
``Yeah, I'm asking if conversations and the interaction with the taskbar is protected and quite private. Like when you are going for therapy, you wouldn't want people outside to hear the things that you say to your therapist. It's a personal thing.''---P4
\end{quote}

Together, these reflections suggest that privacy in AI-mediated support is not just about data protection---it is about control, context, and conditional trust. Anonymity enabled emotional openness for many, but the limits of trust surfaced when participants encountered moments requiring deeper, more personalized engagement. 
Moreover, the diversity of participants' views on data privacy underscores the need for greater attention to privacy-sensitive design in future AI systems.

%% file: 5discussion_new.tex
\section{Discussion}\label{sec:discussion}

\edit{This study explored AD/ADRD caregivers' early perceptions, expectations, and concerns when interacting with Carey in a scenario-driven interview setting. Through these guided interactions, we examined how caregivers make sense of the potential role of an AI chatbot in supporting their informational and emotional needs. Building on our empirical themes identified, we now discuss how our work bears formative design implications, theoretical insights into human–AI support relationships, and clinical considerations for caregiver-facing technologies. Our findings suggest that tools such as \carey{} can be framed as a complementary resource rather than a substitute for human relationships. Participants described the need to turn to an AI-based agent in moments when they lacked immediate access to friends, family, or experts, or when they felt reluctant to burden others with their emotional distress. These reflections suggest that \carey{}'s perceived value lies in its immediacy, low social cost, and judgment-free availability during constrained moments of caregiving.}

\edit{In particular, we situate our findings within prior HCI, digital health, and caregiving research, and reflect on their implications for the design and deployment of caregiver-facing AI systems. We begin by translating our empirical themes into formative design implications (\autoref{sec:design_implications}), directly mapping caregiver needs, perceived system strengths, and unmet gaps to actionable design considerations. We then extend this discussion to broader theoretical and clinical implications (\autoref{sec:theoretical_implications}), considering what these findings reveal about human–AI support relationships and the role of AI in caregiver mental health contexts. 
Finally, we consider ethical implications (\autoref{sec:ethical_implications}), including issues of trust, over-reliance, privacy, and safety that arise when deploying AI systems in sensitive caregiving.}

\begin{table}[t!]
\centering
\sffamily
\footnotesize
\caption{Mapping caregivers' needs and \carey{}'s strengths, gaps, and design recommendations summarizing future explorations where the HCI/design community may have the biggest impact in developing such AI tools.}\label{tab:carey_alignment}
\setlength{\tabcolsep}{3pt}
\begin{tabular}{p{0.11\columnwidth} p{0.22\columnwidth} p{0.23\columnwidth} p{0.3\columnwidth} p{0.1\columnwidth}}
% \toprule
\textbf{Caregivers' Needs} & \textbf{\carey{}'s Strengths} & \textbf{\carey{}'s Gaps} & \textbf{Design Recommendations} & \textbf{\edit{Sections}} \\
\toprule
\edit{On-Demand Information Access} &
\tabitem On-demand access felt practical\newline 
\tabitem Responses were tailored to \hfill \break caregiving scenarios &
\tabitem Trust and source verifiability \hfill \break concerns\newline 
\tabitem Repetitive or generic responses\newline 
\tabitem Limited information verification and source attribution. &
    \tabitem Enhance trustworthiness through \hfill \break transparent source citation and \hfill \break expert verification.\newline \tabitem Adapt responses to caregiving context\newline 
    \tabitem Reduce redundancy, provide more novel \hfill \break (but safe) insights. &  Sections~\ref{sec:result_align_1}, \ref{sec:dis_1}\\
\hdashline
\rowcollight  
\edit{\hfill \break
Safe Space for Disclosure}  &
\tabitem Offered a safe and judgment-free space for candid self-disclosure \hfill \break and self-reflection\newline 
\tabitem Helped reduce stigma around \hfill \break expressing frustration, guilt, \hfill \break and exhaustion \newline 
\tabitem Enabled caregivers to plan \hfill \break difficult decisions &
\tabitem One-sided interactions\newline 
\tabitem Lack of conversational initiative &
\tabitem
Design for conversational turn-taking and add proactive follow-up questions.\newline 
\tabitem Support stigma-free emotional \hfill \break expression through emotionally \hfill \break aware prompts.\newline 
\tabitem Balance ethical objectivity with \hfill \break empathy for caregiving dilemmas. \newline 
\tabitem Facilitate AI-mediated \hfill \break interactive personal journaling that \hfill \break leverages evidence-based \hfill \break psychoeducation. &  Sections~\ref{sec:result_align_2},  \ref{sec:dis_2}\\
\hdashline
\edit{Emotional Support} &
\tabitem Validated complex emotions\newline 
\tabitem Demonstrated empathy and \hfill \break emotionally attuned responses. &
\tabitem Support seemed superficial and in isolated interactions\newline 
\tabitem Lack of dynamic feedback\newline 
\tabitem Fell short of human-human \hfill \break supportive interactions &
\tabitem Model sustained emotional engagement\newline \tabitem Support empathetic interaction loops\newline
\tabitem Design tools to combine and complement AI- and human- driven support.\newline \tabitem Explore scaling up evidence-based therapies (e.g., CBT, DBT) using chatbots. &  Sections~\ref{sec:result_align_3},  \ref{sec:dis_3}\\
\hdashline
\rowcollight \edit{Crisis Management} &
\tabitem 24/7 availability and no wait times in times of need.\newline 
\tabitem Felt emotionally stabilizing. &
\tabitem No crisis escalation mechanism\newline 
\tabitem Risk of false assurance in \hfill \break emergencies. &
\tabitem Integrate crisis detection features\newline \tabitem Enable tiered response systems with clear escalation pathways.\newline \tabitem Design tools for triaging crises and \hfill \break routing to professional interventions &  Sections~\ref{sec:result_align_4}, \ref{sec:dis_4}\\
\hdashline
\edit{Personalization Capabilities} &
\tabitem Responses felt personalized and well-adapted to queries.\newline 
\tabitem Use of personal tone and \hfill \break language (e.g., first-person \hfill \break pronouns) &
\tabitem Only surface-level adaptation\newline 
\tabitem Not adapted to complex \hfill \break caregiving situations &
\tabitem Personalize chatbot interactions based on long-term, evolving caregiving needs. \newline \tabitem Position AI as a complement to human \hfill \break support rather than a replacement.\newline \tabitem Better integrate with real-world \hfill \break resources and referral mechanisms to \hfill \break support communities. &  Sections~\ref{sec:result_align_5},  \ref{sec:dis_5}\\
\hdashline
\rowcollight   
\edit{Privacy and Data Security} &
\tabitem Perceived as non-intrusive.\newline 
\tabitem Did not prompt for unsolicited personal information\newline 
\tabitem Felt safer to share with an AI than a human &
\tabitem Unclear data handling policies\newline 
\tabitem Hesitation to share identifiable information, especially about care recipients\newline 
\tabitem Tension between desire for \hfill \break personalization and privacy \hfill \break concerns &
\tabitem Clearly communicate data usage and \hfill \break retention policies\newline \tabitem Enable user-controlled data sharing/storage.\newline \tabitem Include opt-in and ephemeral \hfill \break data-sharing mechanisms.\newline \tabitem Support pseudonymous use and \hfill \break boundary-setting features.\newline \tabitem Practices to enhance digital and AI \hfill \break literacy and awareness. & Sections~\ref{sec:result_align_6}, \ref{sec:dis_6} \\
\end{tabular}
\end{table}

\subsection{\edit{Implications for} Designing Human-Centered AI to Support AD/ADRD Caregiver Wellbeing} \label{sec:design_implications}

\edit{Building on caregivers' reflections during the interview sessions, we outline early implications for designing human-centered AI systems that support AD/ADRD caregiver wellbeing, as surfaced through scenario-driven interactions with \carey{}.}
\edit{These implications reflect caregivers’ perceptions and expectations of how conversational AI could support their wellbeing, rather than outcomes derived from sustained or real-world use.}

Grounded in established frameworks such as Pearlin's Stress Process Model~\cite{pearlin1981stress} and the Caregiver Appraisal Scale~\cite{struchen2002multidimensional}, we highlight the importance of accounting for the evolving and multifaceted nature of caregiver burden over time. 
Designing for the complexity of caregiving requires AI tools that are not merely responsive, but relational and clinically attuned---sensitive to the emotional, cognitive, and systemic stressors caregivers face~\cite{wu2024social,li2024finding}. These include decision fatigue from navigating fragmented care systems, anticipatory grief during cognitive decline, and role captivity---the sense of being trapped in caregiving responsibilities---all of which are central components of caregiver stress trajectories~\cite{kern2024care, chaudhry2025caregiving,cheung2018anticipatory}. 
To be truly supportive, AI tools must engage not only with the logistical content of caregiving but also with the lived psychological toll it imposes.

\edit{Our analysis suggests that caregivers perceived AI chatbots as potentially addressing certain unmet needs, while also identifying important gaps and tensions that inform future design.}
For each theme identified in~\autoref{sec:result_align}, we map caregiver needs, \carey{}'s strengths and gaps, as well as design recommendations to advance human-centered AI for AD/ADRD caregivers' mental wellbeing. In this section, we systematically map: (1) specific caregiver needs, (2) Carey's demonstrated strengths, (3) current limitations and gaps, and (4) evidence-based design recommendations for future systems. We summarize this information in~\autoref{tab:carey_alignment} and describe these recommendations below:

\subsubsection{Transparent and credible information delivery, integrated with reliable sources of information}
\label{sec:dis_1}
\edit{Caregivers valued \carey{}'s immediate availability but raised concerns about information credibility (\autoref{sec:result_align_1}).}
\edit{Unlike many general mental health contexts where emotional validation alone may suffice, AD/ADRD caregivers may seek evidence-based guidance for complex, high-stakes decisions that affect both their wellbeing and the safety of the care recipient. These needs frequently arise around clinical touchpoints, such as post-diagnosis counseling or care plan changes.}
\edit{A core implication is that caregiver-facing AI should make credibility visible at the point of interaction through traceable sourcing, clear explanations of information provenance, and explicit uncertainty markers when appropriate.}
\edit{Integrating reputable caregiving resources and prioritizing context-aware retrieval can help caregivers assess both the reliability and relevance of information~\cite{yoo2024patient,bhat2023we}.}
\edit{This implication builds on prior HCI and CSCW research emphasizing transparency and trust as foundational for the design of caregiving support tools~\cite{bosch2019caregiver,miller2016partners,seo2019balancing,lee2023reimagining}. Future caregiver-facing AI can draw on explainability research of how the framing and presentation~\cite{van2023measurements}, as well as explicitly conveying uncertainty~\cite{kim2024m} can help shape user trust and reliance.}

\subsubsection{AI-mediated interactive personal journaling} 
\label{sec:dis_2}
\edit{As we found in~\autoref{sec:result_align_3}, caregivers expressed a desire for a safe and judgment-free space for disclosure, and viewed \carey{} as a judgment-free outlet for emotional expressions.}
Caregivers highlighted the value of being able to articulate difficult thoughts without the fear of burdening family members or peers, which aligns with prior evidence that AD/ADRD caregivers experience elevated depression, anxiety, and complicated grief~\cite{puga2022daily}.
\edit{Similar expectations around judgment-free, emotionally supportive spaces have been documented in prior work on parent-facing health technologies and chatbots, where caregivers value opportunities for private reflection, reassurance, and nonjudgmental guidance when managing child health concerns~\cite{felix2024understanding,liebetrau2025engagement}. These parallels suggest that AI-mediated journaling and reflective support can play a meaningful role across caregiving contexts. At the same time, AD/ADRD caregiving introduces distinct challenges, as journaling may need to support prolonged uncertainty, anticipatory grief, and cumulative emotional burden over years rather than discrete developmental phases.}
% {Building on caregivers' desire for a safe space for disclosure described in~\autoref{sec:result_align_3}, \carey{} could potentially serve as a judgment-free outlet for emotional expression.} Caregivers highlighted the value of being able to articulate difficult thoughts without fear of burdening family members or peers, consistent with prior evidence of elevated depression, anxiety, and complicated grief among AD/ADRD caregivers~\cite{puga2022daily}.
This suggests opportunities for AI-mediated journaling as a low-barrier wellbeing practice that complements, rather than replaces, structured interventions. Incorporating optional prompts grounded in CBT, ACT, or DBT principles may support emotional regulation and coping in clinically appropriate ways~\cite{beck1976cognitive,hayes1999act}. At the same time, caregivers’ feedback underscores the importance of avoiding superficial or one-sided interactions, suggesting that journaling features should scaffold reflective depth while preserving psychological safety~\cite{kim2024mindfuldiary}.

\subsubsection{Deepen\edit{ing} emotional engagement beyond surface-level empathy}
\label{sec:dis_3}
% \edit{Addressing the gap between appreciated validation and desired depth in ~\autoref{sec:result_align_3}, 
% \edit{Caregivers caregivers recognized \carey{}'s supportive tone but found interactions fragmented (\autoref{sec:result_align_3}).}
\edit{While caregivers appreciated \carey{}'s emotionally supportive tone, many found the interactions to be fragmented or overly generic (\autoref{sec:result_align_3})}. Emotional validation without continuity can fall short in addressing the long-term psychological needs of caregivers navigating chronic stress and social isolation. 
\edit{This highlights a human-centered design challenge: supporting emotional attunement while maintaining appropriate boundaries. Design implications include consent-driven memory or lightweight continuity mechanisms that reduce the burden of re-explaining context while preserving caregiver control. At the same time, caregivers emphasized that AI should complement—rather than replace—human support, reinforcing the importance of designs that encourage healthy support ecosystems and normalize help-seeking without fostering over-reliance.}

% Designing for deeper emotional connection requires dynamic empathy loops—where the AI can reference prior interactions, recognize emotional patterns over time, and respond with consistency and sensitivity. For instance, P14 noted that they had to explicitly instruct the system to ``talk like a human,'' suggesting the need for algorithms capable of proactively detecting distress through psycholinguistic cues and adapting tone or strategy accordingly. 
% Future tools can implement memory features---combined with user-controlled consent---to reduce conversational management burden, and make the interactions feel more relational and human-like.
% However, our findings also emphasized that \textit{AI should complement---not replace---human support}. 
% While some, like P1, valued the emotional safety of sharing with a nonjudgmental AI, others like P15 cautioned against over-reliance.
% Therefore, technology designers must be attuned to this tension, when building tools for caregivers' wellbeing self-management.

% Our scenario-based study design revealed future technologies can involve scenario-based guidance that helps caregivers navigate through challenging decisions, such as transitioning to professional care or managing difficult behavioral symptoms.
% By integrating case-based reasoning with healthcare ethical principles, AI could provide more sophisticated support for the moral dilemmas caregivers frequently described.
% This approach would preserve psychological safety while helping caregivers navigate moral dilemmas with contextualized, nonjudgmental support.

\subsubsection{Prepar\edit{ing} for crises with tiered intervention mechanisms}
\label{sec:dis_4}
% \edit{Responding to caregivers' crisis management needs in ~\autoref{sec:result_align_4}, the 24/7 availability was valued, but the absence of escalation pathways raised concerns---as P9 shared, while immediate answers were appreciated, the system lacked contextual awareness during emergencies.}

\edit{Caregivers valued \carey{}'s 24/7 availability, but also expressed concerns with the absence of escalation pathways and contextual awareness (\autoref{sec:result_align_4}). AD/ADRD caregiving is often crisis-adjacent, involving situations such as aggression or behavioral escalation~\cite{brodaty2009family}, where inappropriate guidance can have serious consequences.}
\edit{These reflections point to the need for tiered intervention mechanisms that distinguish routine informational or emotional support from potential crisis contexts. 
At a design level, such distinctions could be informed by signals such as explicit expressions of distress, abrupt shifts in emotional tone, or references to loss of control or harm~\cite{sharma2024facilitating,saha2020causal}.
Rather than attempting to resolve crises, AI tools should clearly communicate boundaries, recognize escalation cues, and guide caregivers toward appropriate human or clinical resources. This aligns with HCI work on safety-oriented and ethically constrained conversational health technologies~\cite{slovak2023designing,siddiqui2023exploring}. These tools can implement transparent and caregiver-controlled data practices to enhance trustworthiness in high-stakes scenarios.}

\subsubsection{Personaliz\edit{ing} for the long-term caregiving journey}
\label{sec:dis_5}
% \edit{Addressing the personalization paradox in ~\autoref{sec:result_align_5}, caregivers appreciated surface-level adaptation but sought deeper, journey-based support.}
\edit{We found that caregivers appreciated surface-level adaptation but sought support that evolves across the caregiving journey (\autoref{sec:result_align_5}).}
Caregiving is not a static task, but rather an evolving process marked by shifting emotional, informational, and logistical demands. 
\edit{This suggests designing personalization as a consent-driven, longitudinally aware process that adapts to caregivers’ changing needs. Approaches such as stage-sensitive resources, persistent but optional memory, and integration with existing caregiving infrastructures can reduce repetition and improve relevance without intrusive data collection. Tiered data-sharing models can help balance personalization with privacy, allowing caregivers to calibrate depth of support based on comfort and context.}

\subsubsection{Designing flexible privacy features for conditional trust.}
\label{sec:dis_6}
\edit{Navigating the privacy–personalization tradeoff identified in~\autoref{sec:result_align_6}, caregivers described privacy preferences as conditional and context-dependent.}
\edit{For example, participants described feeling more comfortable disclosing emotional distress when interactions felt anonymous, but expressed hesitation when imagining persistent storage or reuse of sensitive family-related information.}
% \edit{While anonymity sometimes enabled freer disclosure, concerns arose when interactions involved family members, sensitive clinical information, or uncertainty about data use. 
\edit{Design implications include flexible privacy controls such as opt-in data sharing, ephemeral memory, redaction tools, and clear disclosure warnings that allow caregivers to maintain control. Because many AD/ADRD caregivers may have limited familiarity with AI technologies, privacy features must be understandable and accessible, supported by clear, jargon-free explanations and lightweight AI literacy aids. Designing for conditional trust, therefore, requires not only technical safeguards but empathetic and transparent interaction design.}
Drawing on the success of online communities such as Reddit~\cite{de2014mental,saha2017stress,zhou2022veteran,andalibi2016understanding,pickett2024carevirtue}, AI chatbots can facilitate pseudonymous interactions and boundary-setting tools (e.g., disclosure warnings, redaction options) to help reinforce a sense of control and safety.

\subsection{Theoretical and Clinical Implications for Caregiver-Focused Digital Mental Health}\label{sec:theoretical_implications}
Our research builds on and contributes to theoretical understandings within HCI, digital mental health, and caregiving research by foregrounding the affective and relational dimensions of AI-mediated support in AD/ADRD caregiving contexts.
Extending prior work in HCI which explored task-oriented caregiving support (e.g., scheduling, remote monitoring~\cite{arthanat2020caregiver, werner2022helping}), our study reveals how AI systems can also act as \textit{relational agents}---mediating emotional expression, reflection, and companionship. 
This reinforces emerging theories that position AI not merely as a tool but as an individual's ``companion'' to vent out and navigate through decision-making~\cite{foong2024designing,smriti2024emotion}.

Our work also supports the conceptual shift toward emotionally attuned technologies that center emotional labor as integral to user experience and design. 
Rather than treating emotions as secondary to caregiving logistics, our study reveals how technologies like \carey{} can scaffold caregivers' emotional processing around guilt, grief, isolation, and burnout, as similarly noted in prior HCI and chatbot literature~\cite{wawrziczny2017spouse, smriti2024emotion}. This aligns with broader efforts to reconceptualize caregiving as a dynamic, socially embedded, and emotionally demanding experience.

This work contributes to theoretical models of epistemic trust and risk perception in health communication~\cite{kizilcec2016much,jones2024measuring}. Unlike conventional users, 
% of wellbeing apps, 
AD/ADRD caregivers often interpret chatbot responses through the lens of high-stakes responsibility for another person's health. 
Their desire for accurate, evidence-backed, and context-aware responses highlights the need to theorize AI credibility and trustworthiness as situational, emotional, and ethical problems---particularly in domains like caregiving, where inaccurate and misleading information may carry negative consequences for both the caregiver and the care-recipient~\cite{bender2021dangers, inkster2018empathy}.
These insights suggest that emotional and informational trust in a technology are deeply intertwined in caregiving contexts, especially when emotional vulnerability and clinical uncertainty intersect.

From a clinical science perspective, our findings raise important considerations for how AI-mediated journaling and conversational agents might be thoughtfully integrated into mental health care ecosystems for caregivers. 
Although \carey{} was not designed to serve as a therapeutic tool, participants frequently reported emotional relief, catharsis, and a sense of being heard, which are hallmarks of early engagement in emotional processing. 
This suggests that AI tools may offer low-barrier, adjunctive supports to structured interventions like caregiver CBT, psychoeducation, or grief counseling, especially in under-resourced settings~\cite{vaingankar2013perceived, griffiths2016problems,brodaty2009family}.

These dynamics can be interpreted through the Stress Process Model~\cite{pearlin1981stress} and the Caregiver Appraisal Model~\cite{lawton1991two}.
Both of these models emphasize how caregivers' subjective appraisals---such as perceived role captivity, relational strain, and self-efficacy---shape emotional outcomes. 
In this light, AI tool such as \carey{} may help surface and regulate these appraisals in real-time, offering a digital outlet for emotional self-expression that supports coping and adaptation.
Furthermore, \carey{}'s journaling features could serve as early warning systems by identifying patterns associated with distress, grief severity, or burnout. When integrated into clinician dashboards, these features could facilitate timely referrals or adjustments in care planning---though doing so responsibly would require robust ethical safeguards around data use and consent.

Finally, our findings invite interdisciplinary dialogue about the role of AI in emotional meaning-making within caregiving contexts. 
\edit{Across scenarios, caregivers did not engage with \carey{} solely as an information source; instead, they used the interaction to articulate distress, reflect on emotionally complex situations, and make sense of ongoing relational challenges.} \edit{In this way, AI tools like \carey{} can function as companions that help caregivers organize and narrate their experiences, positioning the system as a \textit{co-constructor} of emotional narratives rather than a passive conduit of advice.}
Taken together, our study contributes to a more nuanced theoretical account of how AI can support not only logistical but also emotional and relational aspects of caregiving. \edit{By grounding these insights in caregivers' lived interactions,} our work highlights the value of conceptualizing AI as a mediator of affective experience and motivates closer collaboration across HCI, digital mental health, clinical psychology, and aging studies to better understand the evolving role of relational technologies in caregiver wellbeing.

% Rather than merely acting as information conduits, AI tools like \carey{} may function as \textit{co-constructors} of emotional narratives. This reframing encourages collaboration across clinical psychology, HCI, and implementation science to better align emerging technologies with standards of ethical and effective care.
% Taken together, our study contributes to a more nuanced theoretical account of how AI can support not only logistical but also emotional and relational aspects of caregiving. It underscores the value of conceptualizing AI as a mediator of affective experience and inspires further interdisciplinary theorizing across HCI, digital mental health, clinical psychology, and aging studies to better understand the evolving role of relational technologies in caregiver wellbeing.

\subsection{Ethical Implications}\label{sec:ethical_implications}
% \melissa{look at the CSCW paper}

% \koustuv{This content can be folded into the ethical and privacy implications: However, these potentials must be weighed against the risks of therapeutic drift---where users begin to treat emotionally supportive AI as a substitute for clinical care. To mitigate this, AI systems should include clear disclaimers, escalation pathways, and integration with referral networks to ensure they are positioned as complementary rather than curative supports. Maintaining this distinction is crucial to safeguarding caregiver safety while maximizing the technology's supportive capacity.}

% \koustuv{add about harms and biases}

As AI becomes more integrated into caregiving support, our study surfaces several critical ethical implications. Caregivers expressed concerns about data security and privacy, particularly given their desire for personalized support---creating a fundamental tension where increased personalization requires sharing more sensitive personal information, aligning with prior work on personalization-privacy tradeoff~\cite{asthana2024know, li2024human, zargham2022want}. 
We also caution against interpreting participants' comfort with \carey{} as a blanket endorsement of AI technologies. 
Their trust was shaped within a researcher-guided setting, where the perceived risks were low, and interactions were somewhat bounded. In real-world or commercial applications---where technologies may request more personal information for personalization or even for basic functionalities---additional privacy concerns may arise. These contexts demand greater transparency, user consent, and ethical data practices beyond what our study captured~\cite{chowdhary2023can}. 

% The use of generative AI in caregiving tools raises questions about reliability and potential biases, as AI systems may not fully comprehend the nuances of clinical conditions or caregiving contexts, leading to inappropriate recommendations that could harm vulnerable populations. 

The use of generative AI in caregiving tools also raises concerns about potential bias, reliability, and harm.
AI systems are likely to reflect or amplify existing biases~\cite{ birhane2021multimodal,bender2021dangers}, and may not fully comprehend the nuances of clinical conditions or caregiving contexts, leading to inappropriate recommendations that could harm both the caregivers and care-recipients. 
These harms must be weighed against the risks of therapeutic drift---where users begin to treat emotionally supportive AI as a substitute for clinical care. 
To prevent and mitigate these harms, AI should include clear disclaimers, escalation pathways, and integration with referral networks to ensure they are positioned as complementary rather than curative supports---drawing on the body of work on AI ethics~\cite{raji2022fallacy,amershi2019guidelines}. 
Effective oversight mechanisms should include clinical review boards, caregiver feedback systems, and regular assessment of user outcomes to ensure AI tools maintain appropriate therapeutic boundaries. Maintaining this distinction is crucial to safeguarding caregiver safety while maximizing the technology's benefits.

Additionally, there is a risk that AI and technological solutions may widen existing digital inequalities between those who have access and those who do not~\cite{robinson2015digital}, necessitating efforts to ensure that technologies are inclusive and accessible for caregiving. Our study also highlights the ethical dilemma of AI over-reliance---while AI can supplement caregiving challenges and provide valuable emotional support, it cannot replace the empathy and nuanced understanding that human interaction offers. Therefore, striking a balance between AI automation and human support is crucial to maintaining the dignity and emotional wellbeing of both caregivers and care recipients, ensuring that AI serves as a complement to, rather than the replacement for, human-centered care.

%% file: 6limitations.tex
\subsection{Limitations and Future Directions}
Our work has limitations, which also suggest interesting future directions. 
First, the caregiver participants in our study are not a representative sample. 
However, our primary goal is not to obtain generalizable findings, but to motivate the design and development of AI tools centered around caregivers' needs and expectations.
Then, our study design consisted of scenario-driven interactions with \carey{}---a technology probe designed to elicit user perceptions rather than serve as a high-fidelity prototype or be evaluated for effectiveness.
As such, we did not evaluate \carey{}'s long-term usability or sustained impact.
\edit{Accordingly, the findings should be interpreted as capturing caregivers’ early perspectives and expectations, rather than evidence of real-world adoption, behavioral change, or clinical efficacy.}
Our study inspires the design of future tools, as well as future research on longitudinal deployment studies to assess the feasibility, benefits, and user experience of these AI chatbots in long-term settings. 
The design of such deployments should also employ and evaluate natural interaction patterns and the integration of these tools into caregivers' daily routines.

Further, our work did not measure changes in caregiver burden, psychological distress, or other clinical outcomes before and after interaction with the chatbot. 
Future research should incorporate validated outcome measures (e.g., PHQ-9 for depression~\cite{hivuw_phq9}, Zarit Burden Interview~\cite{wai_zarit} for caregiver burden) to rigorously evaluate the mental health benefits of AI-based support.
% how such AI chatbots perform in real-world caregiving settings, with particular attention to natural interaction patterns and integration into caregivers' daily routines
Additionally, future research can adopt multi-stakeholder approaches that involve not only caregivers, but also clinicians, community health workers, family members, and local support networks. Such inclusive studies can surface diverse perspectives on caregiving needs, assess implementation barriers, and promote the design of AI tools that are contextually appropriate, clinically relevant, and socially acceptable across different caregiving ecosystems.
% eliciting participants' thoughts and perceptions in a scenario-driven technology probe (\carey{})-based interactions. 
% However, \carey---in its current form---was not explicitly designed as a high-fidelity prototype to be evaluated for long-term effectiveness. 
% Along similar lines, our work did not assess changes in caregiver burden, psychological distress, or other clinical outcomes pre/post-chatbot use. 
% Future research should incorporate validated outcome measures (e.g., PHQ-9 for depression~\cite{hivuw_phq9}, Zarit Burden Interview~\cite{wai_zarit} for caregiver burden) to evaluate the efficacy of AI interactions for caregiver mental health. 
Finally, while our findings highlight the potential of conversational AI to support caregiver wellbeing, current limitations underscore the need for more integrated approaches. 
Future feasibility trials should explore pairing AI journaling or check-in tools with clinician dashboards or caregiver support staff. Such hybrid systems may enable real-time responsiveness and continuity of care---key elements in evaluating both the acceptability and clinical effectiveness of these tools in practice. 

%% file: 7conclusion.tex
\section{Conclusion}
% \koustuv{Conclusion needs to be modified.}

In this work, we explored how family caregivers of individuals with Alzheimer's Disease and Related Dementias (AD/ADRD) engaged with \carey{}---a GPT-4o–based AI chatbot---designed to provide mental health support to caregivers. 
Through scenario-driven interactions and semi-structured interviews with 16 caregivers, we examined their needs, expectations, and perceptions of AI-based support. 
Our reflexive thematic analysis revealed five core areas of caregiver expectations---\textit{on-demand information access, emotional support, safe space for disclosure, crisis management}, and \textit{personalization}---each accompanied by nuanced tensions between perceived benefits and concerns. 
While caregivers appreciated Carey's accessibility, nonjudgmental tone, and responsiveness, they also expressed concerns around trust, initiative, emotional over-reliance, and the limits of personalization.
We contributed a systemic understanding of how caregivers interpret AI support, alongside a mapping of their needs, current chatbot strengths, unmet gaps, and design recommendations. 
Some of the major design implications involve providing transparent source citation, conversational turn-taking, interactive personal journaling, tiered response systems for crisis support, alongside balancing personalization with privacy-sensitive features.
Our findings underscore the importance of designing caregiver-centered AI systems that are context-sensitive, emotionally attuned, and capable of evolving with users’ changing needs. This work highlights the potential and limitations of generative AI-driven chatbots in caregiver mental health support and calls for future efforts that center relational, trustworthy, and ethically grounded design.

%% file: 9appendix.tex
% \clearpage
\newpage
\section{Appendix}\label{sec:appendix}
\setcounter{table}{0}
\setcounter{figure}{0}
\renewcommand{\thetable}{A\arabic{table}}
\renewcommand{\thefigure}{A\arabic{figure}}

% edit the codebook with new codes
% compare findings with codebook and probably match any differences for completeness

\begin{table}[h]
\centering
\sffamily
\footnotesize
% \caption{Mean Responses for Each Question (Numerical)}
\caption{Summary of participants' responses to Rapid-Caregivers' Well-being Scale (R-CWBS)~\cite{tebb2013caregiver}. Each question was rated on a scale from: 1 (Rarely), 2 (Occasionally), 3 (Sometimes), 4 (Frequently), and 5 (Usually).}
\label{table:cwbs}

\begin{tabular}{lrrc}
\textbf{Question} & \textbf{Mean} & \textbf{Std. Dev.} & \textbf{Distribution}\\
\toprule
\rowcollight \multicolumn{4}{c}{\textbf{Activities}}\\
Taking care of personal daily activities (meals, hygiene, laundry) & 4.25 & 0.77 & \includegraphics[height=6pt]{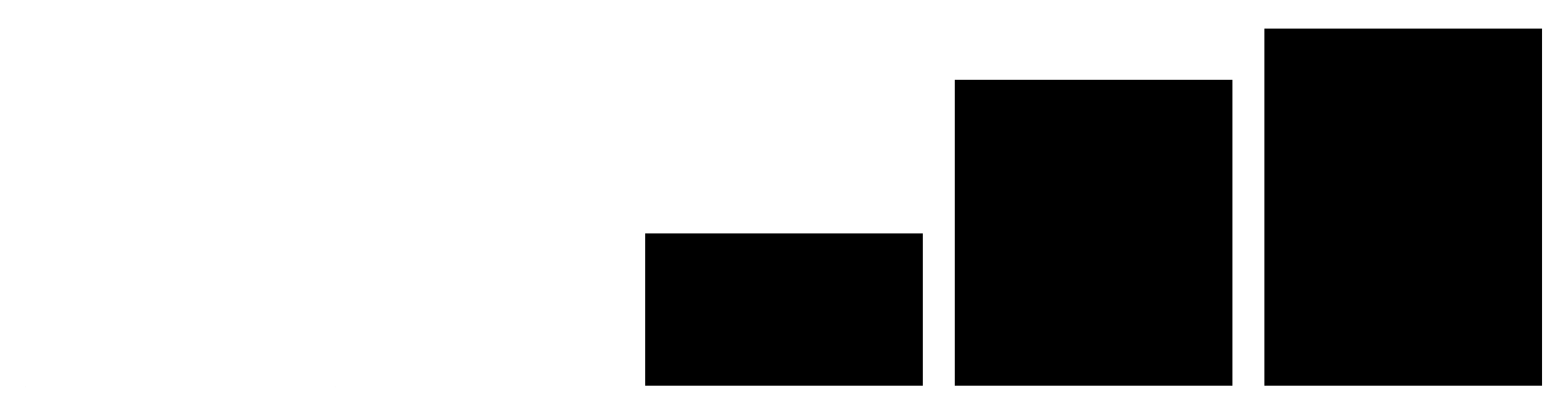}\\
Taking time to have fun with friends and/or family & 3.00 & 1.46 & \includegraphics[height=6pt]{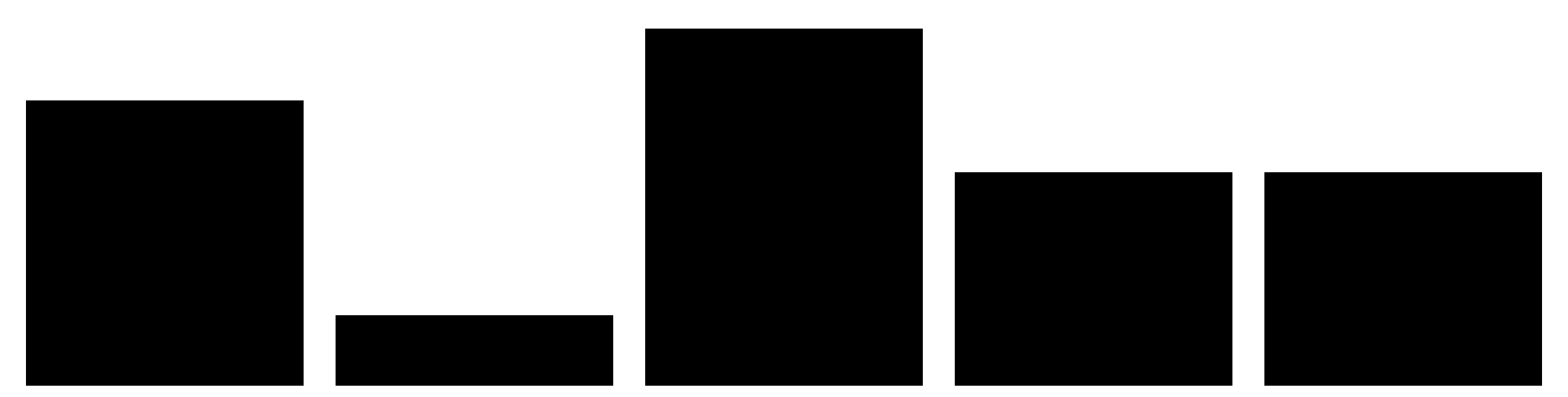}\\
Treating or rewarding yourself & 2.75 & 1.29 & \includegraphics[height=6pt]{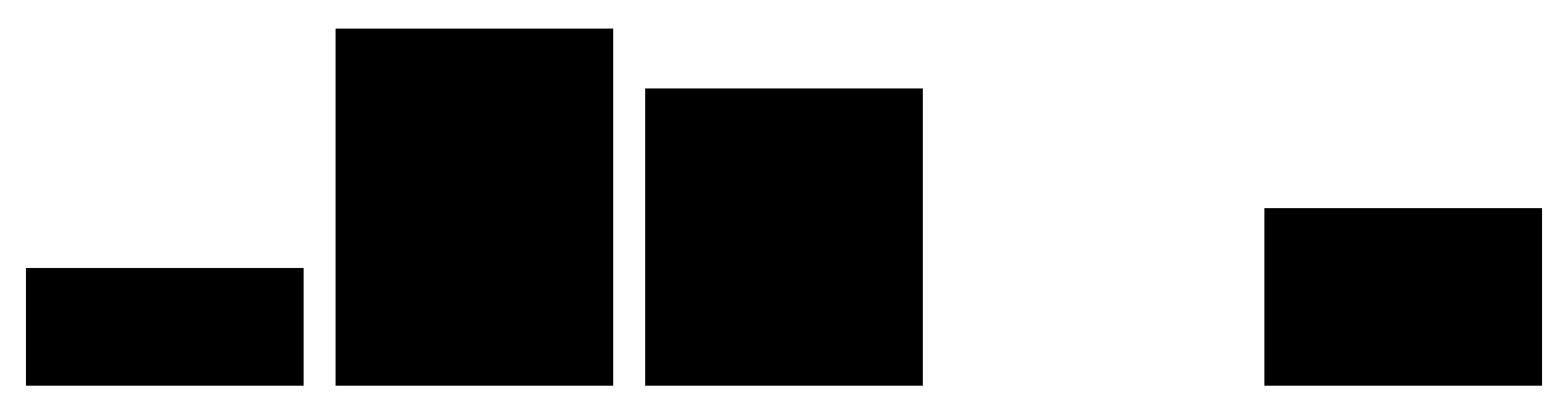}\\
\rowcollight \multicolumn{4}{c}{\textbf{Needs}}\\
Receiving appropriate health care & 3.75 & 1.18 & \includegraphics[height=6pt]{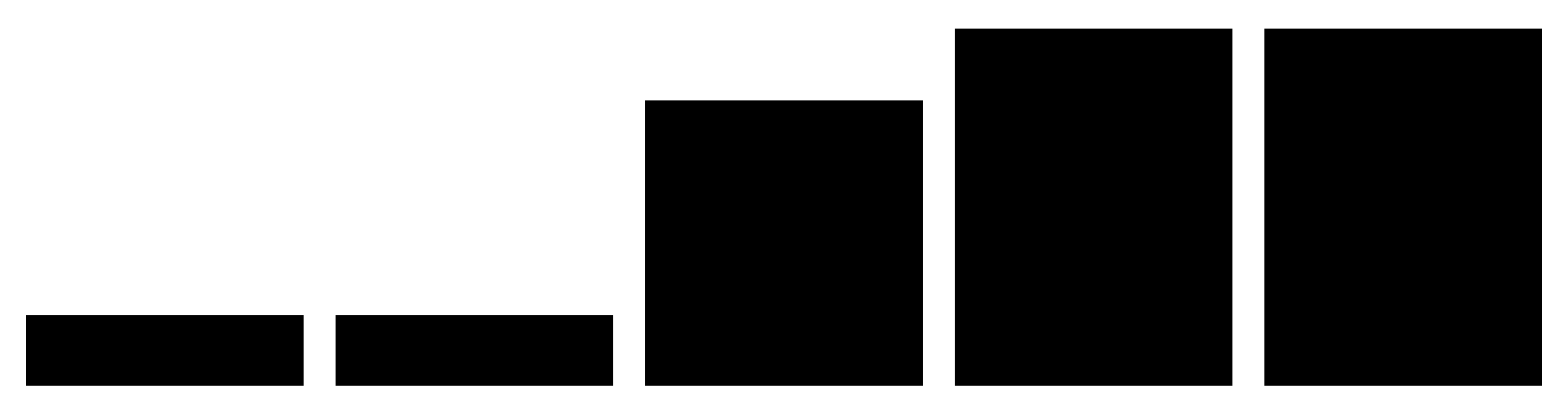}\\
Feeling good about yourself & 3.44 & 1.41 & \includegraphics[height=6pt]{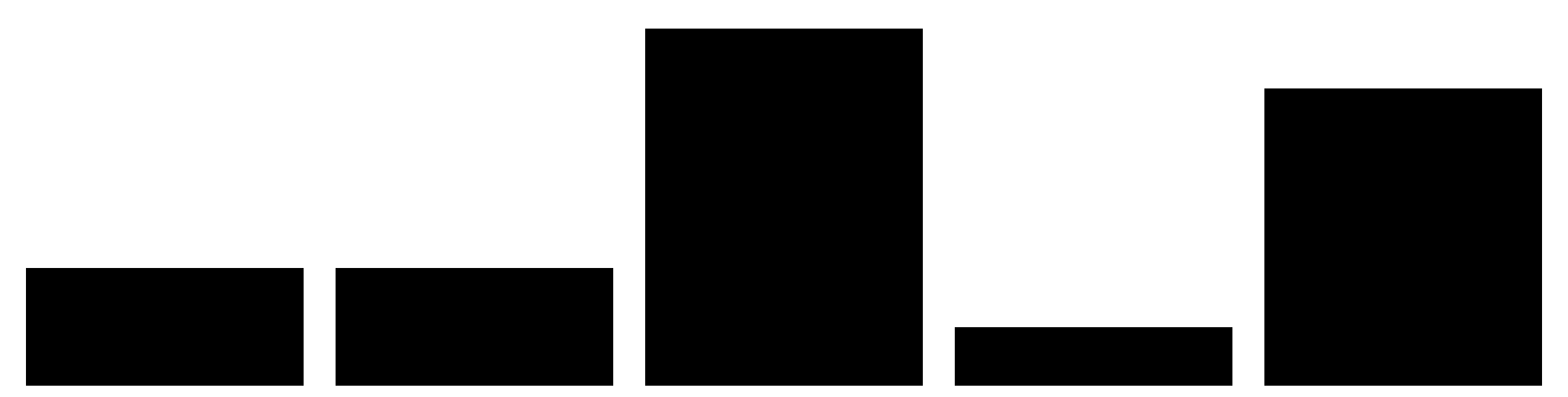}\\
Feeling secure about your financial future & 3.19 & 1.44 & \includegraphics[height=6pt]{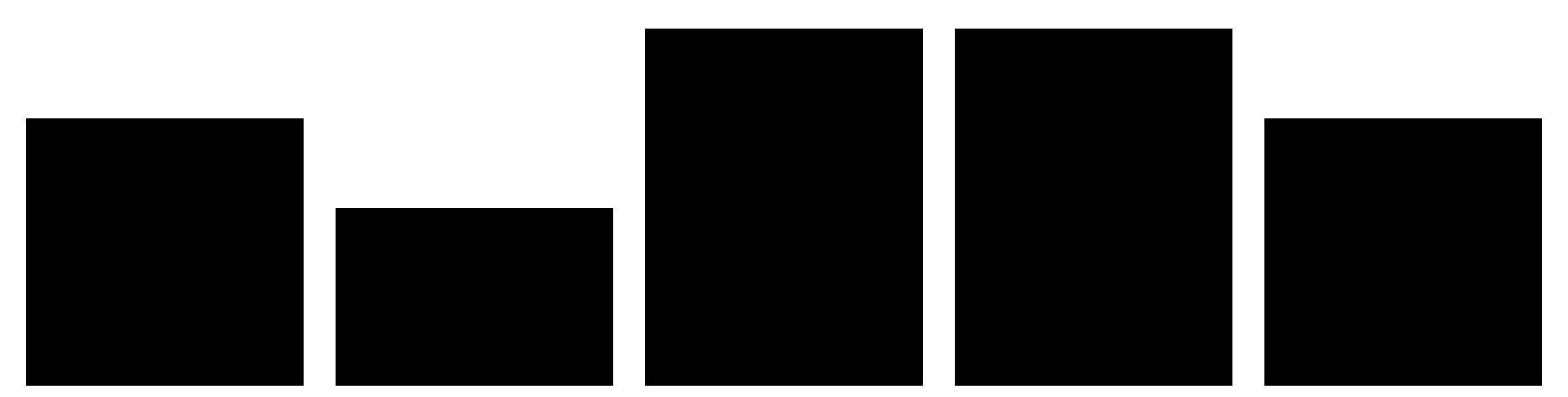}\\
% \bottomrule
\end{tabular}

\end{table}

\begin{table}[h]
\centering
\sffamily
\footnotesize
% \caption{Mean Responses for Each Question (Numerical)}
\caption{Summary of participants' responses to prompts on mental wellbeing concerns from literature. They responded to these prompts based on their level of concern, on a scale of 1 (not at all concerning) to 5 (very concerning).}
\label{table:mentalwellbeing_scale}
\begin{tabular}{lrrc}
\textbf{Question} & \textbf{Mean} & \textbf{Std. Dev.} & \textbf{Distribution}\\
\toprule
% \rowcollight \multicolumn{4}{c}{\textbf{Activities}}\\
Disruptive behaviors by care-recipient & 3.19 & 1.42 & \includegraphics[height=8pt]{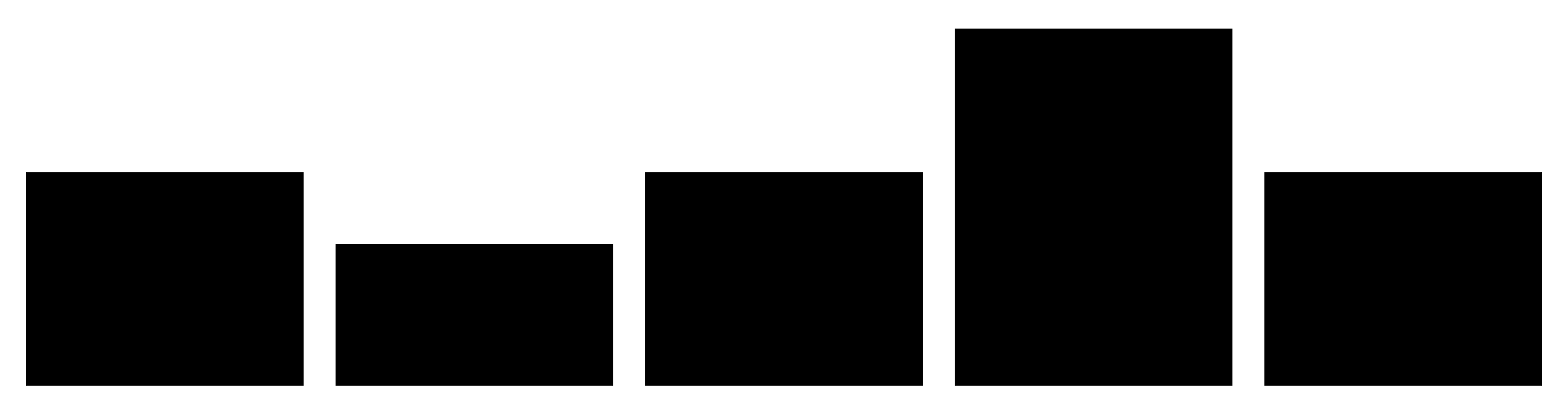}\\
\rowcollight Lack of support & 3.89 & 1.20 & \includegraphics[height=6pt]{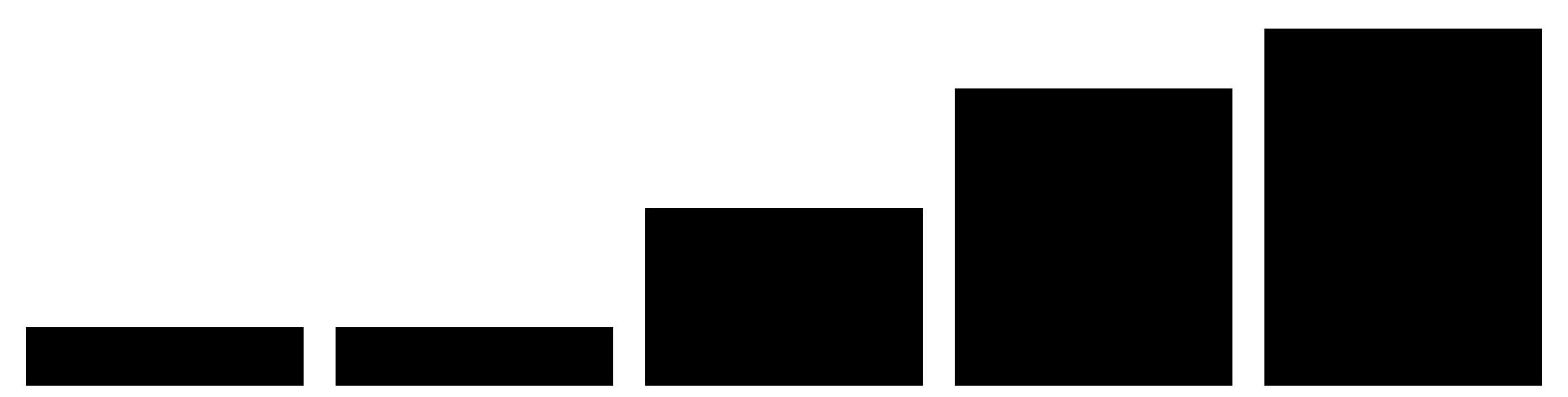}\\
Low self-efficacy & 2.94 & 1.29 & \includegraphics[height=6pt]{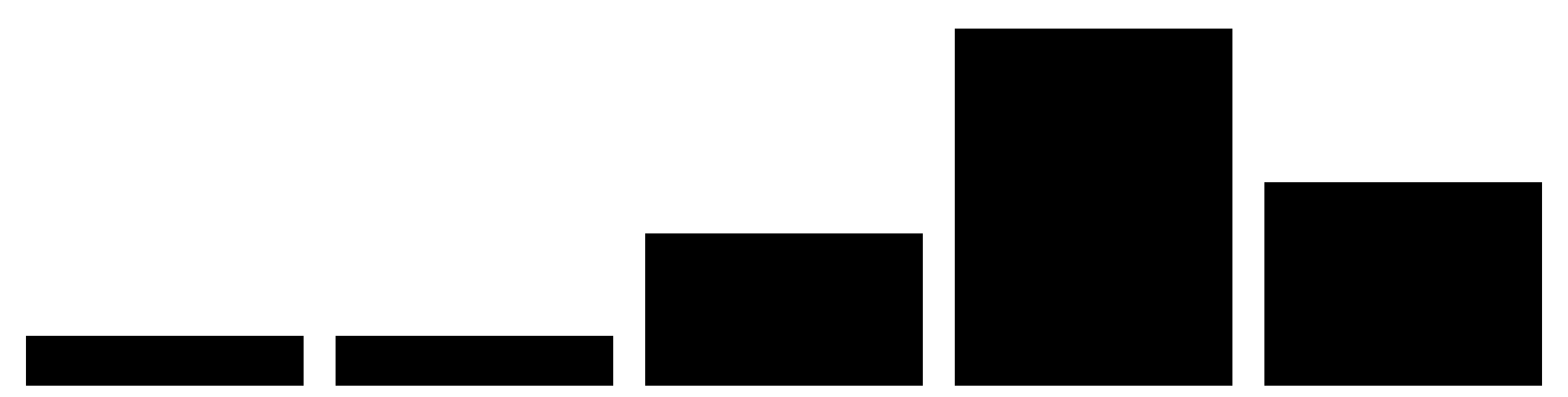}\\
\rowcollight Emotional distress & 3.44 & 1.42 & \includegraphics[height=6pt]{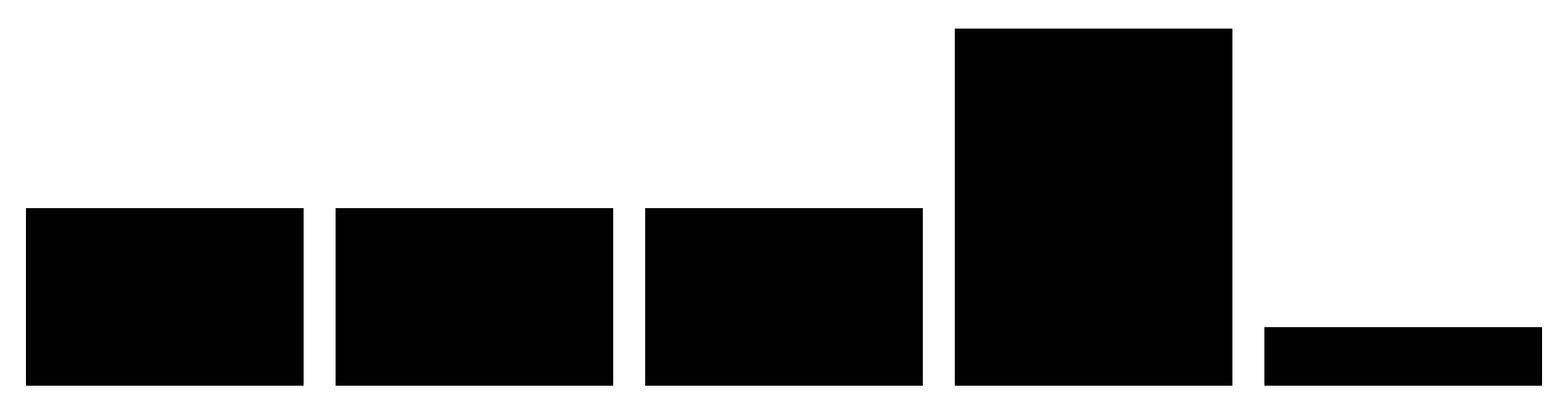}\\
Relationship tensions & 3.38 & 1.20 & \includegraphics[height=6pt]{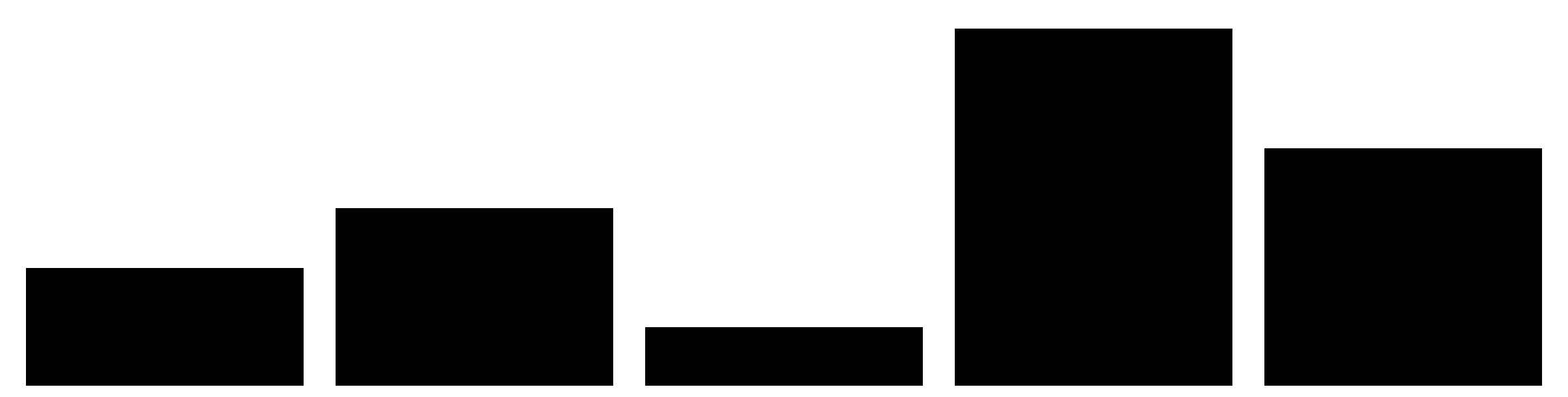}\\
\rowcollight Compassion fatigue & 3.56 & 1.31 & \includegraphics[height=6pt]{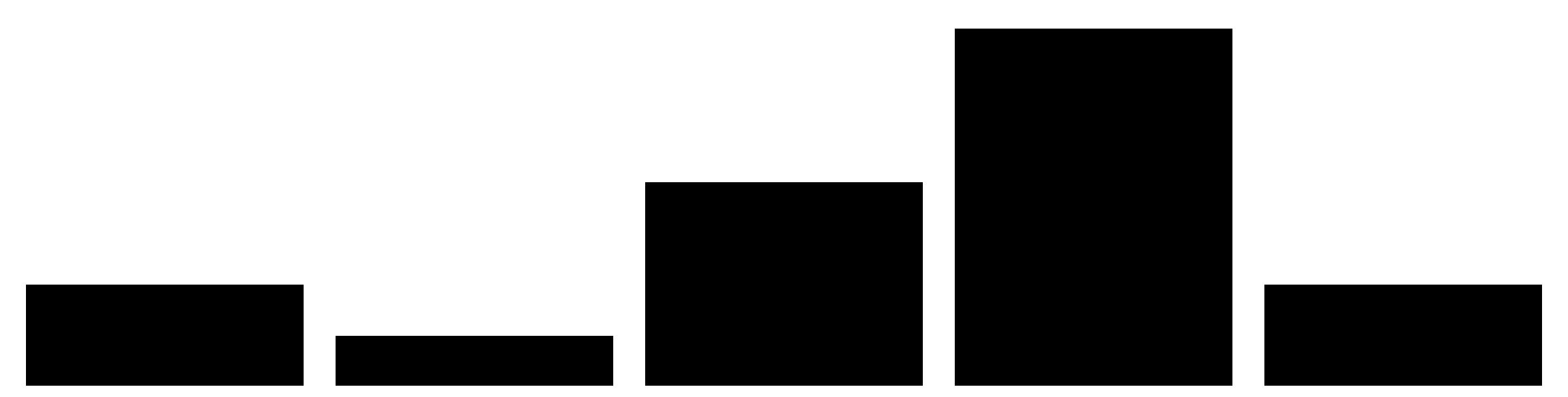}\\
Lack of self-care & 3.56 & 1.46 & \includegraphics[height=6pt]{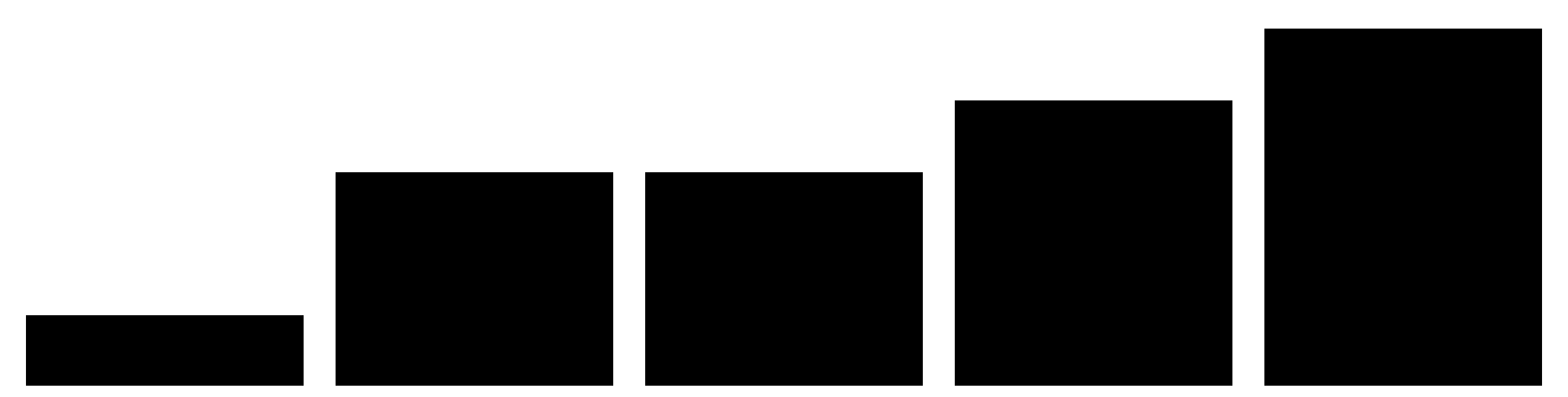}\\
\rowcollight Burnout & 3.75 & 1.00 & \includegraphics[height=6pt]{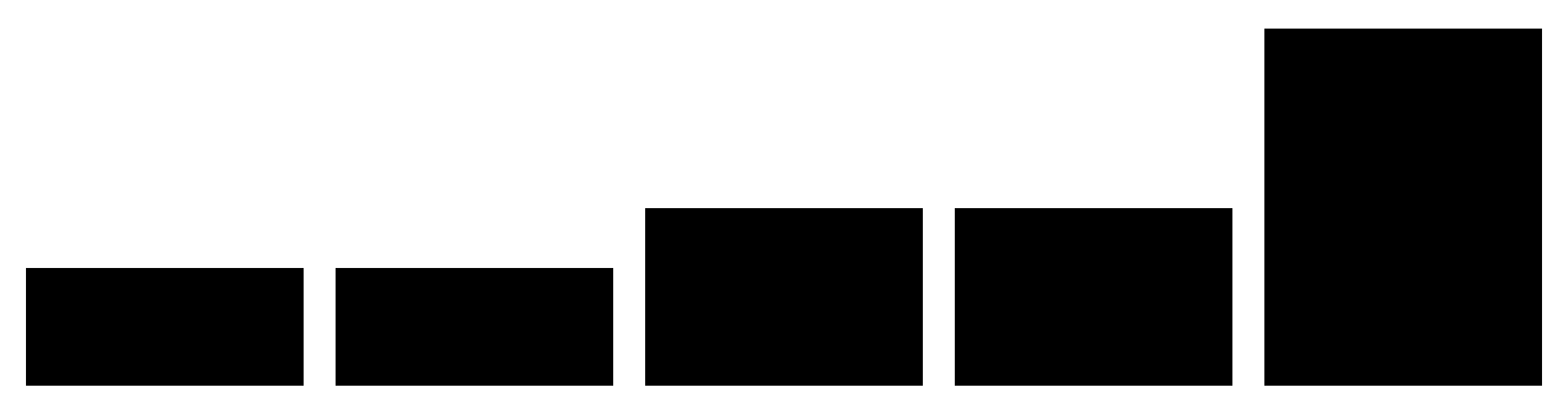}\\
% \bottomrule
\end{tabular}
\end{table}